\documentclass[11pt]{article}
\usepackage{amsmath}
\usepackage{amssymb}

\usepackage{natbib}
\usepackage{graphicx}

\usepackage{color} 

\usepackage[hmargin=2cm, vmargin=2cm]{geometry}
\usepackage[labelfont=bf,labelsep=period, footnotesize]{caption}
\usepackage{sidecap}

\renewcommand\refname{Literature Cited} 

\date{}

\usepackage{cite}

\begin{document}

\begin{flushleft}
{\Huge
\textbf{The evolution of genetic architectures underlying quantitative traits}
}
\bigskip
\\
Etienne Rajon$^{1,2}$, 
Joshua B. Plotkin$^{1}$
\\
\bigskip
\bigskip
$^1$ Department of Biology, University of Pennsylvania, Philadelphia, PA 19104, USA
\\
$^2$ E-mail: rajon@sas.upenn.edu\\
\end{flushleft}

\vspace{1cm}

\begin{abstract} In the classic view introduced by R.A. Fisher, a quantitative
trait is encoded by many loci with small, additive effects. Recent advances in QTL
mapping have begun to elucidate the genetic architectures underlying vast numbers
of phenotypes across diverse taxa, producing observations that sometimes contrast
with Fisher's blueprint. 
Despite these considerable empirical efforts to map the genetic
determinants of traits, it remains poorly understood how the genetic architecture
of a trait should evolve, or how it depends on the selection pressures on the trait.  Here
we develop a simple, population-genetic model for the evolution of genetic
architectures. Our model predicts that traits under moderate selection should be
encoded by many loci with highly variable effects, whereas traits under either
weak or strong selection should be encoded by relatively few loci. We compare
these theoretical predictions to qualitative trends in the genetics of human
traits, and to systematic data on the genetics of gene expression levels in yeast.
Our analysis provides an evolutionary explanation for broad empirical patterns in
the genetic basis of traits, and it introduces a single framework that unifies the
diversity of observed genetic architectures, ranging from Mendelian to Fisherian.
\end{abstract}

\vspace{1cm}

A quantitative trait is encoded by a set of genetic loci whose alleles contribute directly the trait value, interact epistatically to
modulate each others' contributions, and possibly contribute to other traits. The resulting genetic architecture of a trait  \citep{hansen_2006} influences its variational properties \citep{kroymann_mitchell_2005, carlborg_etal_2006, rockman_kruglyak_2006, mackay_etal_2009} and
therefore affects a population's capacity to adapt to new environmental conditions \citep{jones_etal_2004, carter_etal_2005, hansen_2006}. Over longer
timescales, genetic architectures of traits have important consequences for the evolution of recombination \citep{azevedo_etal_2006}, of
sex \citep{devisser_elena_2007} and even reproductive isolation and speciation \citep{fierst_hansen_2010}.

Although scientists have studied the genetic basis of phenotypic variation for more than a century, recent technologies, as well as the promise of
agricultural and medical applications, have stimulated tremendous efforts to map quantitative trait loci (QTL) in diverse taxa \citep{ungerer_etal_2002, flint_mackay_2009, visscher_2008, manolio_etal_2009, brem_etal_2005, brem_kruglyak_2005, rockman_etal_2010, emilsson_etal_2008,
ehrenreich_etal_2012}. These studies have revealed many traits that seem to rely on Fisherian architectures, with contributions from many loci
 \citep{orr_2005}, whose additive effects are often so small that QTL studies lack power to detect them individually  \citep{brem_kruglyak_2005,
rockman_2012, yang_etal_2010}. Other traits, however, are encoded by a relatively small number of loci -- including the large number of human
phenotypes with known Mendelian inheritance.

The subtle statistical issues of designing and interpreting QTL studies in order to accurately infer the molecular determinants of a trait
are already actively studied \citep{brem_kruglyak_2005, rockman_2012, yang_etal_2010}. Nevertheless, distinct from these
statistical issues of inferences from empirical data, we lack a theoretical framework for forming \textit{a~priori} expectations about the genetic architecture
underlying a trait \citep{rockman_kruglyak_2006, hansen_2006}. For instance, what types of traits should we expect to be monogenic, and what traits
should be highly polygenic? More generally, how does the genetic architecture underlying a trait evolve, and what features of a trait shape the
evolution of its architecture? To address these questions we developed a mathematical model for the evolution of genetic architectures, and we compared
its predictions to a large body of empirical data on quantitative traits.

\section*{Results and Discussion}

\subsection*{Genetic architectures predicted by a population-genetic model}

Our approach to understanding the evolution of genetic architectures combines standard models from quantitative genetics \citep{lande_1976} with the
Wright-Fisher model from population genetics \citep{ewens_2004}. In its simplest version, our model considers a continuous trait whose value, $x$, is
influenced by $L$ loci. Each locus $i$ contributes additively an amount $\alpha_i$, so that the trait value is defined as the mean of the $\alpha_i$
values across contributing loci. This trait definition means that a gene's contribution to a trait is diluted when $L$ is large, which prevents direct selection on gene copy numbers when genes have similar contributions \citep{proulx_phillips_2006, proulx_2012}. We discuss this definition below, along with alternatives such as the sum. The fitness of an individual with trait value $x$ is assumed Gaussian with mean $0$ and standard deviation $\sigma_f$, so that
smaller values of $\sigma_f$ correspond to stronger stabilizing selection on the trait \citep{lande_1976}. Individuals in a population of size $N$
replicate according to their relative fitnesses. Upon replication, an offspring may acquire a point mutation that alters the direct effect of one
locus, $i$, perturbing the value of $\alpha_i$ for the offspring by a normal deviate; or the offspring may experience a duplication or a deletion in a
contributing locus, which changes the number of loci $L$ that control the trait value in that individual (see Methods). Point mutations, duplications,
and deletions occur at rates $\mu$, $r_{dup}$, $r_{del}$, which have comparable magnitudes in nature  \citep[table~S1;][]{lynch_etal_2008, watanabe_etal_2009,
lipinski_etal_2011, vanommen_2005}. Finally, an offspring may also increase the number of loci that contribute to its trait value by
recruitment -- that is, by acquiring a recruitment mutation, with probability $\mu \times r_{rec}$, in some gene that did not previously contribute to
the trait value (see Methods). 

Over successive generations in our model, the genetic architecture underlying the trait -- that is, how many loci contribute to the trait's value, and
the extent of their contributions -- varies among the individuals in the population, and evolves. The genetic architectures that evolve in our model
represent the complete genetic determinants of a trait, which may include -- but do not correspond precisely to -- the genetic loci that would be detected based on polymorphisms
segregating in a sample of individuals in a QTL study. We discuss this important distinction below, when we compare the predictions of our model to
empirical QTL data.
\begin{SCfigure}
\centering
\includegraphics[width=87mm]{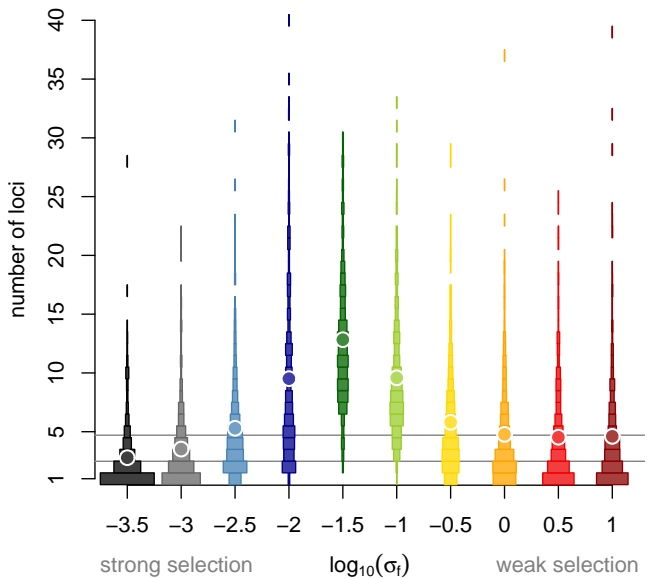}
\caption{The genetic architecture underlying a trait depends on the strength of selection on the trait, in a population-genetic model. Traits subject to intermediate selection (intermediate values of $\sigma_f$) evolve genetic architectures with the greatest number of controlling loci. Dots denote the mean number of loci in the architecture underlying a trait, among $500$ replicate Wright-Fisher simulations, for each value of the selection pressure $\sigma_f$. The rectangular areas represent the distribution of the number of loci in the architecture. The neutral expectations for the equilibrium number of loci (see Methods) are represented as grey lines, when recruitment events are neutral (top line) or not (bottom line). Parameters are set to their default values (table~S2).}
\end{SCfigure}

We studied the evolution of genetic architectures in sets of $500$ replicate populations, simulated by Monte Carlo, with different amounts of selection on the trait. We ran each of these simulations for $50$ million generations, in order to model the extensive evolutionary divergence over which genetic architectures are assembled in nature. The form of the genetic architecture that evolves in our model depends critically on the strength of selection on the trait. In particular, we found a striking non-monotonic pattern: the equilibrium number of loci that influence a trait is greatest when the strength of selection on the trait is intermediate (Fig. 1). Moreover, the variability in the contributions of loci to the trait value (Fig.~S1) and the effects of deleting or duplicating genes (Fig.~S2) are also greatest for a trait under intermediate selection. In other words, our model predicts that traits under moderate selection will be encoded by many loci with highly divergent effects; whereas traits under strong or weak selection will be encoded by relatively few loci. 

We also studied how epistatic interactions among loci influence the evolution of genetic architecture. To incorporate the influence of locus $j$ on the contribution of locus $i$ we introduced epistasis parameters $\beta_{ji}$ so that the trait value is now given by 
\begin{equation}
x=\dfrac{1}{L} \displaystyle \sum_{i=1}^{L} \biggl(\alpha_i \times f_\beta \biggl( \sum_{j=1}^{L} \beta_{ji}\biggr) \biggr), \end{equation}
where $f_\beta$ is a standard sigmoidal filter function \citep[][see Methods and Fig.~S4]{azevedo_etal_2006}. As with the direct effects of loci, the epistatic effects were allowed to mutate and vary within the population, and evolve. Although significant epistatic interactions emerge in the evolved populations (Fig.~S3B), the presence of epistasis does not strongly affect the average number of loci that control a trait (Figs.~S3A and S4). Epistasis is not required for the evolution of large $L$, nor does it change the shape of its dependence on the strength of selection. 

\subsection*{Intuition for the results} 

There is an intuitive explanation for the non-monotonic relationship between the selection pressure on a trait and the number of loci that control it.
For a trait under weak selection (high $\sigma_f$), changes in the trait value have little effect on fitness. Thus, even if deletions, recruitments
and duplications change the trait value, these changes are nearly neutral (Fig.~2). As a result, the number of loci controlling the trait evolves to
its neutral equilibrium, which is small because deletions are more frequent than duplications and recruitments (see Methods, Figs.~1 and S3). On the
other hand, when selection on a trait is very strong (low $\sigma_f$), few point mutations, and only those with small effects on the trait, will fix
in the population. As a result, all loci have similar contributions to the trait value (Fig.~2 -- row 1), and so duplications or deletions again have
little effect on the trait or on fitness (Fig.~2 -- rows~2 and 3). In this case, the equilibrium number of loci is given by the value expected when
deletions and duplications, but not recruitments, are neutral (Figs. 1 and S3). Only when selection on a trait is moderate can variation in the
contributions across loci accrue and impact the fixation of deletions and duplications (Fig.~2 -- row 4), by a process called compensation: a slightly
deleterious point mutation at one locus, which perturbs the trait value, segregates long enough to be compensated by point mutations at other
loci \citep{rokyta_etal_2002, Meer_etal_2010, kimura_1985, poon_otto_2000}. Compensation increases the variance in the contributions among loci (Fig. 2,
row 1), as has been observed for many phenotypes in plants and animals \citep{rieseberg_etal_1999}. Finally, even though duplications and deletions are
mildly deleterious in this regime, there is a bias favoring duplications over deletions (Fig.~2 -- row 3). This bias arises because duplications
increase the number of loci in the architecture, which attenuates the effect of each locus on the trait (Fig.~2 -- row~2). Thus when selection is
moderate, duplications and recruitments fix more often than deletions and drive the number of contributing loci above its neutral expectation (Fig.~2
-- rows 4 and 5). As the number of loci increases the bias is reduced (Fig.~2 -- rows 4 and 5), and so $L$ equilibrates at a predictable value
(Figs.~1 and S3). Duplications and recruitments might also be slightly favored over deletions under intermediate selection, because architectures with
more loci also have reduced genetic variation \citep{wagner_etal_1997}. This effect -- which would positively select for an increase in gene copy
numbers -- is likely weak in our model, as duplications and recruitments are deleterious on average under intermediate selection, only less so than deletions (Fig.~2 -- rows 4 and 5).
\begin{figure}
\centering
\includegraphics[width=180mm]{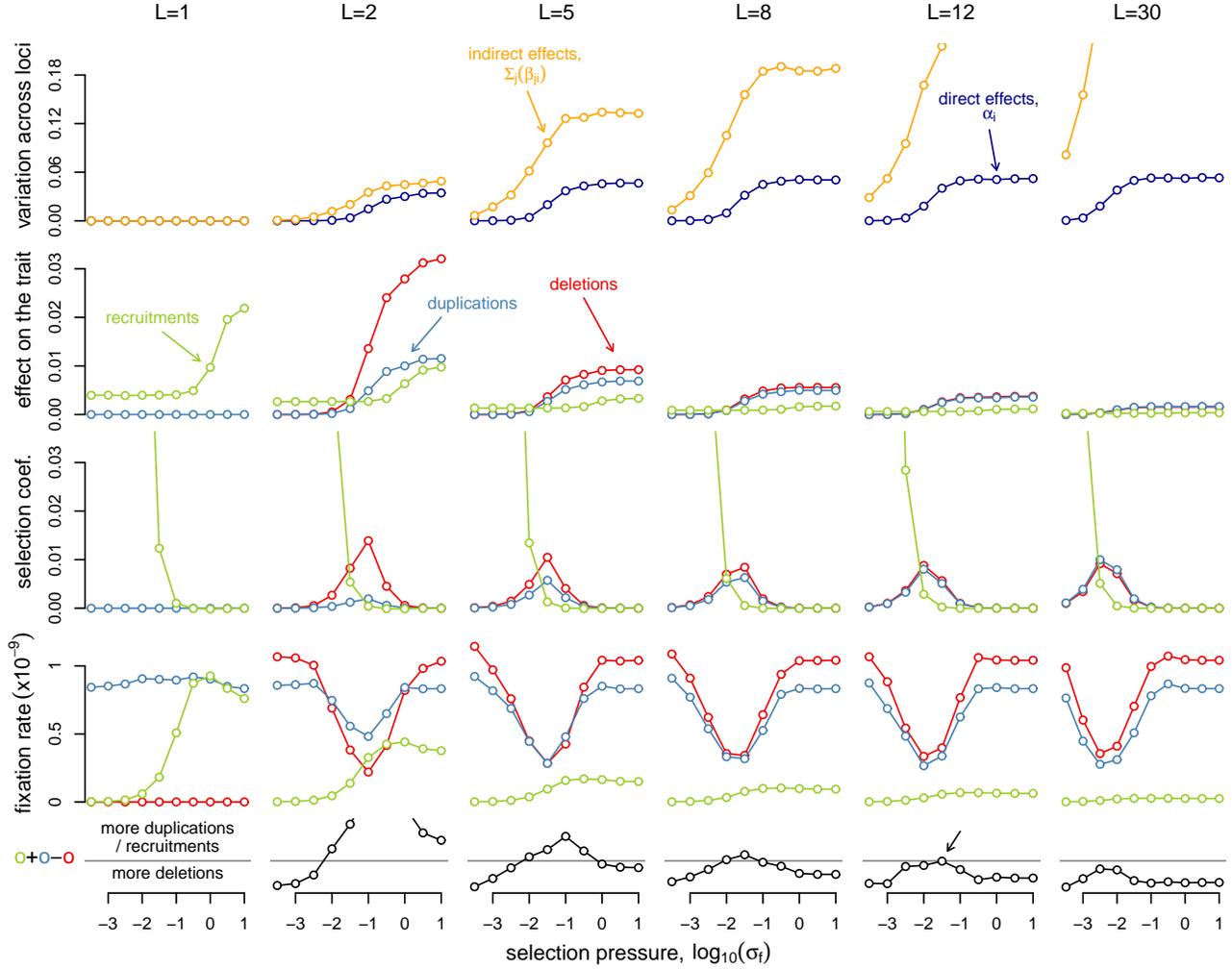}
\caption{The consequences of gene duplications, recruitments and deletions in a population-genetic model. Populations were initially evolved with a fixed number of controlling loci $L$ (line~1), and we then measured the effects of recruitments, deletions and duplications on the trait value (line~2) and on fitness (line~3). From the latter, we calculated the rate at which deletions, recruitment and duplications enter and fix in the population (line~4), and the resulting rate of change in the number of loci contributing to the trait (line~5). \textbf{Line~1:} For $L>1$, the variation in direct effects ($\alpha_i$) and indirect effects among controlling loci ($\sum_j (\beta_{ji})$) increases as selection on the trait is relaxed. \textbf{Line~2:} As a consequence of this variation among loci, the average change in the trait value following a duplication or a deletion also increases as selection on the trait is relaxed. \textbf{Line~3:} Changes in the trait value are not directly proportional to fitness costs, because the same change in $x$ has milder fitness consequences when selection is weaker (larger $\sigma_f$). As a result, the average fitness detriment of duplications and deletions is highest for traits under intermediate selection. \textbf{Line~4:} Consequently, the fixation rates of duplications and deletions are smallest under intermediate selection. \textbf{Line~5:} The equilibrium number of loci controlling a trait under a given strength of selection is determined by that value of $L$ for which duplications and recruitments on one side, and deletions on the other, enter and fix in the population at the same rate. For example, when $\sigma_f=10^{-1.5}$ these rates are equal when $L$ is close to $12$ (black arrow), so that the equilibrium genetic architecture contains $\approx 12$ loci on average (compare Fig. S3 black arrow).}
\end{figure}

\subsection*{Robustness of results to model assumptions}

The predictions of our model -- notably, that the number of loci in a genetic architecture is greatest for traits under intermediate selection -- are
robust to choices of population-genetic parameters. The non-monotonic relation between selection pressure on a trait and the size of its genetic
architecture, $L$, holds regardless of population size; but the location of maximum $L$ is shifted towards weaker selection in larger populations
(Fig. S5).  This result is compatible with our explanation involving compensatory evolution: selection is more efficient in large populations, and so
compensatory evolution occurs at smaller selection coefficients. Likewise, when the mutation rate is smaller the resulting equilibrium number of
controlling loci is reduced (Fig. S6). This result is again compatible with the explanation of compensatory evolution, which requires frequent
mutations. Increasing the rate of deletions relative to duplications also reduces the equilibrium number of loci in the genetic architecture, but our
qualitative results are not affected even when $r_{del}$ is twice as large as $r_{dup}$ (Fig. S7). Finally, increasing the rate of recruitment
$r_{rec}$ (or the genome size) increases the number of loci contributing to all traits except those under very strong selection, as expected from Fig.
2. Our prediction that traits under intermediate selection are encoded by the richest genetic architectures is insensitive to changes in this
parameter, and it holds even in the absence of recruitment (Fig. S8).

Our analysis has relied on several quantitative-genetic assumptions, which can be relaxed. First, we assumed that all effects of locus $i$
(\textit{i.e.} $\alpha_i$ and all $\beta_{ij}$ and $\beta_{ji}$) are simultaneously perturbed by a point mutation. Relaxing this assumption, so that
a subset of the effects are perturbed, does not change our results qualitatively (Fig. S9).  Second, we assumed that point mutations have unbounded
effects so that variation across loci can increase indefinitely. To relax this assumption we made mutations less perturbative to loci with large
effects (see Methods). Even a strong mutation bias of this type led to very small changes in the equilibrium behavior (Fig. S10). Third, we assumed
no metabolic cost of additional loci, even though additional genes in \textit{Saccharomyces cerevisiae} are known to decrease fitness
slightly \citep{wagner_2005b, wagner_2007}. Nonetheless, including a metabolic cost proportional to $L$ does not alter our qualitative predictions
(Fig. S11). Finally, we defined the trait value as the average of the contributions $\alpha_i$ across loci, as opposed to their sum. This definition
reflects the intuitive notion that a gene product's contribution to a trait will generally depend on its abundance relative to all other
contributing gene products. Moreover, this assumption that increasing the number of loci influencing a trait attenuates the effect of each one is supported by empirical data: changing a gene's copy number is known to have milder
phenotypic effects when the gene has many duplicates \citep{gu_etal_2003, conant_wagner_2004}. Nonetheless, alternative definitions of the trait
value, which span from the sum to the average of contributions across loci, generically exhibit the same qualitative results (text S1 and Fig. S12).

Although robust to model formulation and parameter values, our results do depend in part on initial conditions. When selection is strong, the initial
genetic architecture can affect the evolutionary dynamics of the number of loci (Fig.~S14). This occurs because the initial architecture may set
dependencies among loci that prevent a reduction of their number. This result indicates that only those architectures of traits under very strong
selection should depend on historical contingencies. We have also studied a multitrait version of our model, where genes participating in other traits
can be recruited or lost through mutation. Even though this model features pleiotropy, and the effects of recruitments evolve neutrally, our
qualitative results remained unaffected (text~S3 and Fig.~S15).

\subsection*{The dynamics of copy number}

Previous models related to genetic architecture have been used to study the evolutionary fate of gene duplicates. These models typically assume that
a gene has several sub-functions, which can be gained \citep[neo-functionalization;][]{ohno_1970} or lost \citep[sub-functionalization;][]{force_etal_1999,
lynch_force_2000} in one of two copies of a gene. Such ``fate-determining mutations'' \citep{innan_kondrashov_2010} stabilize the
two copies, as they make subsequent deletions deleterious. Such models complement our approach, by providing insight into the evolution of discrete,
as opposed to continuous or quantitative, phenotypes. Yet there are several qualitative differences between our analysis and previous studies
of gene duplication. Most important, our model considers the dynamics of both duplications and deletions, in the presence of point mutations that
perturb the contributions of loci to a trait. This co-incidence of timescales is important in the light of empirical data  \citep{lynch_etal_2008,
watanabe_etal_2009, lipinski_etal_2011, vanommen_2005} showing that changes in copy numbers occur at similar rates as point mutations (table~S1).
Under these circumstances, a gene may be deleted or acquire a loss-of-function mutation before a new function is gained or lost. Our model
includes these realistic rates, and accordingly we find that duplicates are very rarely stabilized by subsequent point mutations. Instead, the
number of loci in a genetic architecture may increase, in our model, because compensatory point mutations introduce a bias towards the fixation of
duplications as opposed to deletions. 

\subsection*{Comparison to empirical eQTL data}

Like most evolutionary models, our analysis greatly simplifies the mechanistic
details of how specific traits influence fitness in specific organisms.  As a
result, our analysis 
explains only the broadest, qualitative features of how genetic architectures vary
among phenotypic traits, leaving a large amount of variation unexplained.  This
remaining variation may be partly random (as predicted by the distributions of the
number of evolving loci, see \textit{e.g.} Fig. 1), and partly due to ecological
and developmental details that our model neglects.

Due to this variation, a quantitative comparison between our model and empirical data would require information
about the genetic architectures for at least hundreds of traits (see below, for our analysis
of expression QTLs). 
Nevertheless, the qualitative, non-monotonic predictions of
our model (Fig.~1) may help to explain some well-known trends in the genetics
of human traits.  For instance, in accordance with our predictions, human traits
under moderate selection, such as stature or susceptibility to mid-life diseases
like diabetes, cancer, or heart-disease, are typically complex and highly
polygenic; whereas traits under very strong selection, such as those
(\textit{e.g.} mucus composition or blood clotting) affected by childhood-lethal
disease like Cystic fibrosis or Haemophilias are often Mendelian; and so too traits
under very weak selection (such as handedness, bitter taste, or hitchhiker's
thumb) are often Mendelian. Our analysis provides an evolutionary explanation for
these differences, and it delineates the selective conditions under which we may
expect a Mendelian, as opposed to Fisherian, architecture.

We tested our evolutionary model of genetic architectures by comparison with
empirical data on a large number of traits. Such a comparison must, of course,
account for the fact that our model describes the true genetic architecture
underlying a trait, whereas any QTL study has limited power and describes only the
associations detected from polymorphisms segregrating in a particular sample of
individuals.  Accounting for this discrepancy (see below), we compared our model
to data from the study of \citet{brem_etal_2005}, who measured
mRNA expression levels and genetic markers in $112$ recombinant strains produced
from two divergent lines of \textit{S. cerevisiae}. For each yeast transcript we
computed the number of non-contiguous markers associated with transcript level, at
a given false discovery rate (see Methods). We also calculated the codon
adaptation index (CAI) of each transcript -- an index that correlates with the
gene's wildtype expression level and with its overall importance to cellular
fitness \citep{sharp_li_1987}. We found a striking, non-monotonic relationship
between the CAI of a transcript and the number of loci linked to variation in its
abundance (Fig.~3A). Thus, assuming that CAI correlates with the strength of
selection on a transcript, \citet{brem_etal_2005} detected
more loci regulating yeast transcripts under intermediate selection than
transcripts under either strong or weak selection.

We compared the empirical data on yeast eQTLs (Fig.~3A) to the predictions of our evolutionary model. In order to make this comparison, we first
evolved genetic architectures for traits under various amounts of selection (Fig.~S3), and for each architecture we then simulated a QTL study of the
exact same type and power as the yeast eQTL study: that is, we generated 112 crosses from two divergent lines using the yeast genetic map (text~S2). 
As expected, the simulated QTL studies based these 112 segregants detected many fewer loci linked to a trait than in
fact contribute to the trait in the true, underlying genetic architecture (Fig.~3B versus Fig.~1). This result is consistent with previous
interpretations of empirical eQTL studies \citep{brem_kruglyak_2005}. The simulated QTL studies revealed another important bias: a locus that
contributes to a trait under weak selection is more likely to be correctly identified in a QTL study than a locus that
contributes to a trait under strong selection (Fig.~S16). Furthermore, our simulations demonstrate that the number of associations detected
in such a QTL study depends on the divergence time between the parental strains used to generate recombinant lines (Fig.~S17). Finally, traits under
weaker selection may be more prone to measurement noise, which we also simulated (Fig.~S18). Despite these detection biases, which we have
quantified, the relationship between the selection pressure on a trait and the number of \textit{detected} QTLs in our model (Fig.~3B and
Figs.~S18 and S19) agrees with the relationship observed in the yeast eQTL data (Fig.~3A). Importantly, both of these relationships exhibit
the same qualitative trend: traits under intermediate selection are encoded by the richest genetic architectures.
\begin{figure}[h] \centering \includegraphics[width=180mm]{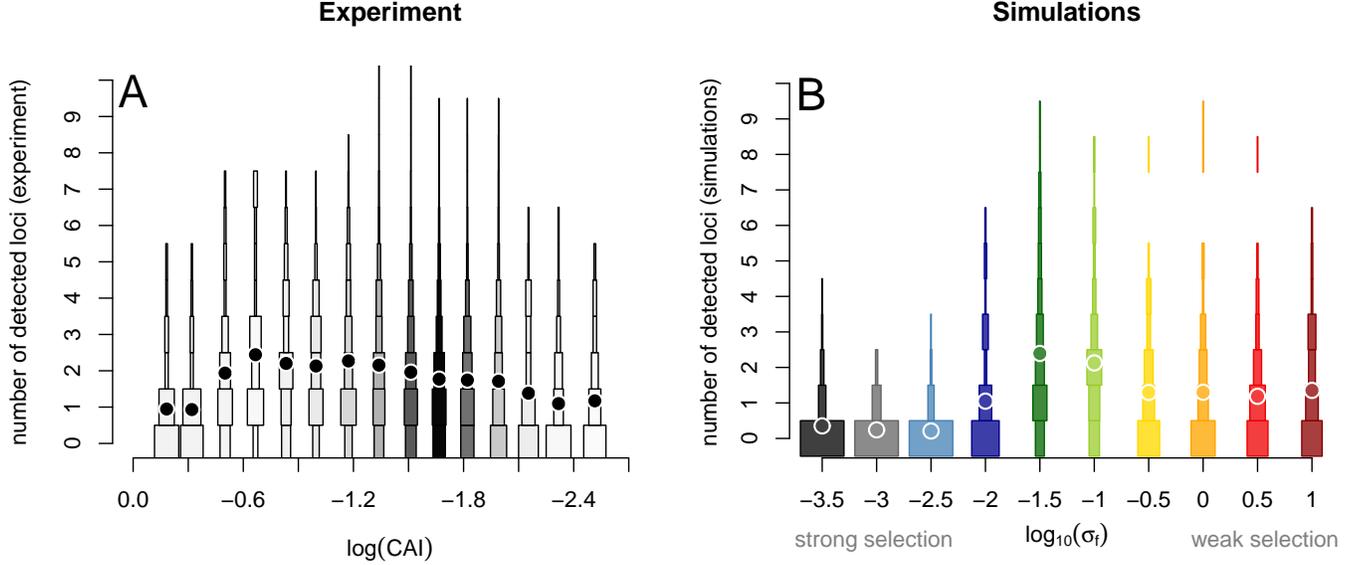} \caption{The
number of genetic loci controlling a trait inferred from real \textit{S.
cerevisiae} populations (panel A) and from simulated populations (B) has a
non-monotonic relationship with the strength of selection on the trait. A: In
the yeast data of Brem \citet{brem_etal_2005}, the
largest number of eQTLs were detected for those transcripts
(\textit{i.e}, traits) under intermediate levels of selection (intermediate CAI),
whereas fewer eQTLs were detected for transcripts under either weak or strong
selection. Transcripts were binned according to their log CAI values. Squares
represent the distribution of the number of one-way eQTLs identified from the
study of \citet{brem_etal_2005}, for traits within each bin of CAI.
Greyscale indicate the number of transcripts in
each bin (darker means more data). Mean numbers of detected eQTLs are represented
by circles. B: For the simulated experiment, we evolved $100$ populations of
genetic architectures, using the parameters corresponding to Fig. S3.
From each such population, we then evolved two lines independently for $25,000$ generations in the
absence of deletions, duplications and recruitment, to mimic the divergent
strains used in the yeast cross of \citet{brem_etal_2005}. From these two
divergent genotypes we then created $112$ recombinant lines following the genetic
map from \citet{brem_etal_2005}. 
We then analyzed the resulting simulated data with \textit{R/qtl}
in the same way as we had analyzed the yeast data (text S2). The distribution of QTLs detected
and their means are represented as in Fig. 1, for each
value of selection strength $\sigma_f$.} \end{figure}

\section*{Conclusion} 

Many interesting developments lie ahead. Our model is far too simple to account for tissue- and time-specific gene expression, dominance, context-dependent effects, etc \citep{mackay_etal_2009, korpela_etal_2011}. How these complexities will change predictions for the evolution of genetic architectures remains an open question. Nonetheless, our analysis shows that it is possible to study the evolution of genetic architecture from first principles, to form \textit{a priori} expectations for the architectures underlying different traits, and to reconcile these theories with the expanding body of QTL studies on molecular, cellular, and organismal phenotypes.

\section*{Methods}

\subsection*{Model} We described the evolution of genetic architectures using the Wright-Fisher model of a replicating population of size $N$, in which haploid individuals are chosen to reproduce each generation according to their relative fitnesses. The fitness of an individual with $L$ loci encoding trait value $x$ is
\begin{equation} 
\omega_k=G(x, 0, \sigma_f) \times (1-L \times c) 
\end{equation}
where $G$ denotes the density at $x$ of a Gaussian distribution with mean $0$ and standard deviation $\sigma_{f}$, and the second term denotes the metabolic cost of harboring $L$ loci, which depends on a parameter $c$. The trait value of such an individual, given the direct contributions $\alpha_i$ and epistatic terms $\beta_{ji}$ is described by Eq. (1) where 
\begin{equation} 
f_\beta(y)=\dfrac{2}{1+e^{-s_\beta y}}
\end{equation} is a sigmoidal curve, so that the epistatic interactions either diminish or augment the direct contribution of locus $i$ depending on whether $\sum_j \beta_{ji}$ is positive or negative (Fig.~S4). In general, loci do not influence themselves ($\beta_{ii} \equiv 0$) and, in the model without epistasis, all $\beta_{ji} \equiv 0$ and $f_\beta \equiv 1$. If an individual chosen to reproduce experiences a duplication at locus $i$ then the new duplicate, labelled $k$, inherits its direct effect ($\alpha_k=\alpha_i$) and all interaction terms ($\beta_{kj}=\beta_{ij}$ and $\beta_{jk}=\beta_{ji}$ for all $j \neq i,k$), with the interaction terms $\beta_{ik}$ and $\beta_{ki}$ initially set to zero. Recruitment occurs with probability $r_{rec}$ per mutation of one of the $6,000$ genes not contributing to the trait. The initial direct contribution $\alpha_i$ of recruited locus $i$ is drawn from a normal distribution with mean zero and standard deviation $\sigma_m$; its interaction terms with other loci ($k$), $\beta_{ik}$ and $\beta_{ki}$, are initially set to zero. Note that this assumption is relaxed in the multilocus version of our model, where the direct and indirect effects of recruitments evolve neutrally (text S3 and Fig. S15). 

In general a point mutation at locus $i$ changes its contribution to the trait, $\alpha_i$, and all its epistatic interactions, $\beta_{ij}$ and
$\beta_{ji}$, each by an independent amount drawn from a normal distribution with mean zero and standard deviation $\sigma_m$. The normal distribution
satisfies the assumptions that small mutations are more frequent than large ones \citep{orr_1999, eyre_keightley_2007}, and that there is no mutation
pressure on the trait \citep{lande_1976}. We relaxed the former assumption by drawing mutational effects from a uniform distribution without
qualitative changes to our results (Fig.~S13). In order to relax the latter assumption we included a bias towards smaller mutations in loci with
large effects, so that the mean effect of a mutation at locus $i$ now equals $-b_\alpha \times \alpha_i$ and $-b_\beta \times \beta_{ij}$,
respectively for $\alpha_i$ and $\beta_{ij}$ \citep{rajon_masel_2011}. We also considered a model in which a mutation at locus $i$ affects only a
proportion $p_{em}$ of the values $\alpha_i$, $\beta_{ij}$, and $\beta_{ji}$. By default, simulations were initialized with $L=1$ and $\alpha_1=0$;
alternative initial conditions were also studied, as shown in Fig. S14.

\subsection*{Markov chain for neutral changes in copy number}
When deletions and duplications are neutral, and recruitments strongly deleterious, the evolution of the number of loci $L$ in the genetic architecture is described by a Markov-chain on the positive integers. The probability of a transition from $L=i$ to $L=i+1$ equals $r_{dup} \times i$, and that of a transition from $i$ to $i-1$ is $r_{del} \times i$. We disallow transitions to $L=0$, assuming that some regulation of the trait is required. We obtained the stationary distribution of $L$ by setting the density of $d_1$ of individuals in stage $1$ to $1$ and calculating the density $d_i$ of individuals in the following stages as

\begin{equation}
	d_i=\dfrac{r_{dup} \times (i-1)}{r_{del} \times i} d_{i-1}
\end{equation}
The equilibrium probability of being in state $i$ was calculated as
\begin{equation}
	p_i=\frac{d_i}{\sum_{i=1}^{\infty} d_i}
\end{equation}
and the expected value of $L$ was calculated as $\sum_{i=1}^{\infty} p_i \times i$. With $r_{dup}=10^{-6}$ and $r_{del}=1.25 \times 10^{-6}$, we found an equilibrium expected $L$ of $2.485$.

When deletions, duplications and recruitments are all neutral, equation (4) can be replaced by:
\begin{equation}
	d_i=\dfrac{r_{dup} \times (i-1) + 6000 \times \mu \times r_{rec}}{r_{del} \times i} d_{i-1}
\end{equation}
This equation illustrates the fact that the rates of deletions (which include loss of function mutations) and duplication depend on the number of loci in the architecture, whereas the rate of recruitments does not. With $\mu=3 \times 10^{-6}$ and $r_{rec}=5 \times 10^{-5}$, we found an equilibrium expected $L$ of $4.705$.

\subsection*{Calculation of $\overline{s}$ and $\overline{p_{\text{fix}}}$}
We first evolved populations to equilibrium with a fixed number of controlling loci $L$, and we then measured the effects of deletions,
duplications or recruitments introduced randomly into the population. We simulated the evolution of the genetic architecture with $L$ fixed in $500$
replicate populations, over $8 \times 10^6$ generations for deletions and $10 \times 10^6$ generations for duplications, reflecting the unequal
waiting time before the two kinds of events. We used $10 \times 10^6$ generations for recruitment as well, although different durations did not affect our results. For each genotype $k$ in each evolved population, we calculated the fitness $\omega_k(i)$ of mutants with locus $i$ deleted or duplicated. We calculated the corresponding selection coefficients as:
\begin{equation}
s_k(i)=\dfrac{\omega_k(i)}{< \omega >}-1
\end{equation}
where $< \omega >$ denotes mean fitness in the population. We calculated $\overline{s}$ as the mean across loci and genotypes of $s_k(i)$, weighted
by the number of individuals with each genotype. We calculated the probability of fixation of a duplication, deletion or recruitment as
\begin{equation}
p_{\text{fix}}(s_k(i))=\dfrac{1-e^{-2 s_k(i)}}{1-e^{-2 N s_k(i)}},
\end{equation}
and obtained the mean $p_{\text{fix}}$ using the same method as for $\overline{s}$.

Rates of deletions and duplications fixing were calculated per locus (Fig. 2) as $r_{del}$ or $r_{dup}$ times $p_{fix}$. The
total probability of a duplication or a deletion entering the population and fixing is, of course, also multiplied by $L$. However, recruitment rates
remain constant as $L$ changes. Therefore, we divided the rate of recruitments by $L$ in Fig. 2, for comparison to the per-locus duplication and
deletion rates.

\subsection*{Number of loci influencing yeast transcript abundance}
We used the \textit{R/qtl} \citep{broman_etal_2003,r_cite} package to calculate LOD scores for a set of $1226$ observed markers and $3223$ uniformly distributed pseudomarkers separated by $2$ cM, by Haley-Knott regression. We calculated the LOD significance threshold for a false discovery rate (FDR) of $0.2$ as the corresponding quantile in the distribution of the maximum LOD after $500$ permutations (a FDR of $0.01$ and a fixed LOD threshold of $3$ produced qualitatively similar results). The number of detected loci linked to the expression of a transcript was calculated as the number of non-consecutive genomic regions with a LOD score above the threshold.  We downloaded \textit{S. cerevisiae} coding sequences from the Ensembl database (EF3 release), and calculated CAI values with the \textit{seqinr} \citep{cite_seqinr} package, using codon weights from a set of $134$ ribosomal genes.

\clearpage
\begin{flushleft}
{\Huge
\textbf{Supplementary material}
}
\end{flushleft}

\renewcommand{\thefigure}{S\arabic{figure}}
\renewcommand{\thetable}{S\arabic{table}}
\renewcommand{\theequation}{S\arabic{equation}}

\section*{Text S1: Alternative definitions of the trait}
As described in the main text, the non-monotonic relationship between the strength of selection on a trait and the number of loci in its
underlying genetic architecture depends on the unequal fitness consequences of deletions and duplications. This behavior should therefore be absent when the trait value $x$ is the sum, rather than the average, of the contributions across loci. To explore this issue, we generalized our definition of the trait by introducing an additional parameter $\epsilon$:
	\begin{equation}
		x=\dfrac{1}{1 + \epsilon (L-1)} \displaystyle \sum_{i=1}^{L} \biggl(\alpha_i \times f_\beta \biggl( \sum_{j=1}^{L} \beta_{ji}\biggr) \biggr).
	\end{equation}
As the parameter $\epsilon$ ranges from 0 to 1 the trait definition ranges from a sum to an average. As expected when $\epsilon=0$, the number of loci in the equilibrium genetic architecture is greatly reduced under intermediate selection as compared to the results in the main text (Fig. \ref{fig:epsilon}). Interestingly, the mean number of loci also shows a non-monotonic trend in this situation, with a small peak at $\log_{10}(\sigma_f)=-0.5$. This trend is likely driven by the fixation rate of recruitment mutations (as seen in Fig. 2). This explains why it is much less pronounced than those in Figs. 1 and S3-A, which involve a higher fixation rate of duplications over deletions. For any other value of $\epsilon<1$, we found a non-monotonic relationship similar to the one reported in the main text. Thus, our qualitative results hold for all models provided the trait is not defined strictly as the sum of contributions across loci.

When the trait value equals the sum of the contributions of all locus ($\epsilon=0$), the effect of a gene deletion, knock-out or knock-down is independent of the number of copies of the gene. 
Conversely when the trait value is the mean of the contributions ($\epsilon=1$), or is some function between the mean and the sum ($0 <
\epsilon < 1$), the effect of a deletion decreases with the number of loci in the genetic architecture. As shown by \citet{conant_wagner_2004} in $C. elegans$, the number 
of detectable knock-down phenotypes decreases 
with the number of copies of genes in a gene family, suggesting that $\epsilon$ does indeed exceed $0$ in this species. A similar stronger effect of the deletion of a singleton compared to that of a duplicate has also been observed in \textit{S. cerevisiae} \citep{gu_etal_2003}. 
	
\section*{Text S2: QTL detection in simulated populations}
We analyzed the
genetic architectures that evolved under our population-genetic model using a simulated QTL study of the exact same type and power as the yeast eQTL
study \citep{brem_etal_2005}. Specifically, $100$ evolved populations were taken from simulations with parameters corresponding to Fig. S2 for the
model with epistasis. From each population, we evolved two lines independently for $T$ generations in the absence of deletions, duplications and
recruitment. We then used the most abundant genotype from each line to create parental strains, mimicking the diverged BY and RM parental strains in \citet{brem_etal_2005}. A few populations were polymorphic for the number of loci initially, sometimes resulting in two lines with different values of $L$, which we discarded. In each parent, we assigned the $L$ contributing loci randomly among $1226$ simulated marker sites, and also assigned their associated $\alpha_i$ values and the interactions $\beta_{ij}$ between loci. We constructed $112$ recombinant haploid offspring by mating these two parents according to the genetic map inferred from Brem et al. 

Each offspring inherited each $\alpha_i$ value, and the set of interactions towards other loci ($\beta_{ij}$ $\forall j$), from either one or the other parent. The trait value in each offspring was calculated as in eq. (1) and then was perturbed by adding a small amount of noise (normally distributed with mean $0$ and standard deviation $\sigma_n$), to simulate measurement noise. We then analyzed these artificial genotype and phenotype data following the same protocol we used for the real yeast eQTLs data (i.e. using Rqtl). We repeated this entire process with $100$ different pairs of parents for each value of $\sigma_f$. 

Fig. S18 shows the relationship between the selection pressure $\sigma_f$ and the number of linked loci detected in this simulated QTL study, for
different divergence times between the two lines and different values of $\sigma_n$. In Fig. S19, we increased $\sigma_n$ proportionally to
$\log_{10}(\sigma_f)$, from $0.0001$ to $0.001$. We also calculated the probability that a locus known to influence the trait in the true
architecture (Fig. S3) is in fact detected in the QTL study. This probability is plotted as a function of $\sigma_f$ for different values of the noise $\sigma_n$ (Fig. S16) and of the time of divergence (Fig. S17).

\section*{Text S3: Multitrait model}
We simulated the evolution of the genetic architecture underlying multiple traits with a model slightly modified from the single-trait version. In this model, the phenotype consists of $10$ traits, each trait $k$ under a different selection pressure $\sigma_f(k)$ (the values of $\sigma_f(k)$ are those used in independent simulations of the single-locus model; see the x-axis of Fig. S15). In the multiple traits version, $L$ denotes the total number of loci forming the architecture of the $10$ traits. $L$ can change when loci are duplicated at rate $r_{dup}$ and deleted at rate $r_{del}$. Each locus participates to a set of traits. The direct effect of locus $i$ on trait $t$ is now denoted $\alpha_{it}$ and the indirect effect of locus $i$ on the part of locus $j$ that contributes to $t$ is denoted $\beta_{ijt}$. 

To allow for partial gains and losses of function, we define two new matrices $A$ and $B$, which have the same dimensions as $\alpha$ and $\beta$. The functions corresponding to $\alpha_{it}$ and $\beta_{ijt}$ are `on' when $A_{it}=1$ or $B_{ijt}=1$, respectively, and are `off' otherwise. Similarly to eq. (1) in our single-trait model, we calculate the trait value $t$ as: 
	\begin{equation}
		x_t=\displaystyle \sum_{i=1}^{L} \biggl(A_{it} \alpha_{it} \times f_\beta \biggl( \sum_{j=1}^{L} B_{jit} \beta_{jit}\biggr) \biggr) / \displaystyle \sum_{i=1}^L A_{it}
	\end{equation}
where $f_\beta$ is the sigmoidal function defined in eq (3). Point mutations of locus $i$ alter all $\alpha_{it}$ and $\beta_{ijt}$ by a normal deviate. Moreover, a mutation can change $A_{it}$ and $B_{ijt}$ to $0$ with probability $0.1$ and to $1$ with probability $0.005$. Over successive generations, the genetic architecture underlying each trait evolves through gene deletions and duplications, and through recruitments and losses of new functions. In this model, only the $L$ genes in the simulated architecture can be recruited -- \textit{i.e.} we do not assume a fixed number of genes that can be recruited at any time. Therefore, the phenotypic effects of recruitment evolve during our simulation, instead of being sampled from a given distribution. 

If $\sum_i A_{it}=0$ for any trait $t$, the individual is considered non-viable and fitness $\omega_k$ equals $0$. Otherwise, fitness is the product of Gaussian functions for each trait times the cost associated to the number of loci, as follows:
\begin{equation} 
\omega_k= \displaystyle \prod_{k=1}^{10} G(x_k, 0, \sigma_f(k)) \times (1-L \times c) 
\end{equation}
We simulated the evolution of the genetic architecture through a Wright Fisher process, with population genetics parameters identical to the default values in table S2, except $c=10^{-4.5}$ \citep{wagner_2005b, wagner_2007}). The results of $200$ simulations are represented in Fig. S15.

\renewcommand\refname{Additional reference}

\clearpage

\begin{table}	
\caption{Estimates of rates of mutations $\mu$, gene duplications $r_{dup}$ and deletions $r_{del}$. All rates are per gene per generation. $\mu$
is the rate of non-silent mutations \citep{lynch_etal_2008} ($0.75 \times$ the per-nucleotide mutation rate). When the mutation rate was given per
nucleotide, we multiplied it by the average gene length in Eukaryotes \citep{xu_etal_2006} ($1346$ bp). For \textit{D.
melanogaster} \citep{watanabe_etal_2009}, the rate of detectable mutations was used, after correcting for the length of the $3$ loci in the study. 
The scale of analysis can be the whole genome (WG), or a specific set of loci, in which case the number of loci is denoted in the table.}
\centering
\begin{tabular}[c]{| l | c | c | c | c | c | r |}
		\hline
		\footnotesize Species & \footnotesize $\mu$ & \footnotesize $r_{dup}$ & \footnotesize $r_{del}$ & \footnotesize Scale & \footnotesize Refs \\ \hline
		\footnotesize \textit{S. cerevisiae} & \footnotesize $3.33 \times 10^{-7}$ & \footnotesize $3.4 \times 10^{-6}$ & \footnotesize $2.1 \times 10^{-6}$ &\footnotesize WG & \footnotesize \citep{lynch_etal_2008} \\
		\footnotesize \textit{D. melanogaster} & \footnotesize $9.18 \times 10^{-7}$ & \footnotesize $4 \times 10^{-7}$ & \footnotesize $4 \times 10^{-7}$ &\footnotesize 3 & \footnotesize \citep{watanabe_etal_2009} \\
		\footnotesize \textit{C. elegans} & \footnotesize $2.02 \times 10^{-6}$ & \footnotesize $1.25 \times 10^{-7}$ & \footnotesize $1.36 \times 10^{-7}$ &\footnotesize WG & \footnotesize \citep{lipinski_etal_2011} \\
		\footnotesize \textit{H. sapiens} & \footnotesize $1.5 \times 10^{-5}$ & \footnotesize $10^{-5}$ & \footnotesize $6.67 \times 10^{-5}$ &\footnotesize 1 & \footnotesize \citep{vanommen_2005} \\
\hline
\end{tabular}	
\label{tab:rates}

\end{table}

\begin{table}	
\caption{Definition of parameters, their default values, and range of values examined in the corresponding figures.}
\begin{center}
\begin{tabular}[c]{| l | p{5cm} | c | p{4cm} | c |}
\hline
		\footnotesize Parameter name & \footnotesize Definition & \footnotesize Default value & \footnotesize Values used & \footnotesize Fig.  \\ \hline
\footnotesize $s_{\beta}$ & \footnotesize Slope of $f_\beta()$ & \footnotesize $4$ & \footnotesize $\{ 1, 4, 16 \}$ & \footnotesize \ref{fig:slope_beta}  \\
\footnotesize $N$ & \footnotesize Population size & \footnotesize $1000$ & \footnotesize $\{ 100, 1000, 10000 \}$ & \footnotesize \ref{fig:N}  \\
\footnotesize $\mu$ & \footnotesize Mutation rate & \footnotesize $3 \times 10^{-6}$ & \footnotesize $\{ 0.33, 1, 3, 6 \} \times 10^{-6}$ & \footnotesize \ref{fig:mu}  \\
\footnotesize $r_{dup}$ & \footnotesize Duplication rate & \footnotesize - & \footnotesize $10^{-6}$ & \footnotesize -  \\
\footnotesize $r_{del}$ & \footnotesize Deletion rate & \footnotesize $1.25 \times 10^{-6}$ & \footnotesize $\{ 1.25, 1.5, 2 \} \times 10^{-6}$ & \footnotesize \ref{fig:r_del}  \\
\footnotesize $r_{rec}$ & \footnotesize Probability of recruitment after the mutation of a non-contributing locus & \footnotesize $5 \times 10^{-5}$ & \footnotesize $\{ 2.5, 5, 10 \} \times 10^{-5}$ & \footnotesize \ref{fig:r_rec}\\
\footnotesize $\sigma_f$ & \footnotesize SD of the fitness function & \footnotesize $[ 10^{-3.5} - 10 ]$ & \footnotesize - & \footnotesize All  \\
\footnotesize $\sigma_m$ & \footnotesize SD of mutation effect function & \footnotesize $0.01$ & \footnotesize - & \footnotesize -  \\
\footnotesize $p_{em}$ & \footnotesize Probability that a subfunction is changed by a mutation & \footnotesize $1$ & \footnotesize $\{ 0.25, 0.5, 1 \}$ & \footnotesize \ref{fig:p_em}  \\
\footnotesize $b_{\alpha}$ & \footnotesize Mutation bias on $\alpha$ & \footnotesize $0$ & \footnotesize $\{ 0, 0.2, 0.4 \}$ & \footnotesize \ref{fig:bias}  \\
\footnotesize $b_{\beta}$ & \footnotesize Mutation bias on $\beta$ & \footnotesize $0$ & \footnotesize $\{ 0, 0.2, 0.4 \}$ & \footnotesize \ref{fig:bias}  \\ 
\footnotesize $c_L$ & \footnotesize Metabolic cost of $L$ loci & \footnotesize $0$ & \footnotesize $\{ 0, 10^{-4.5}, 10^{-3.5} \}$ & \footnotesize \ref{fig:c}  \\ \hline
\end{tabular}	
\end{center}
\label{tab:parameters}
\end{table}

\clearpage

		\setkeys{Gin}{width=150mm}
	\begin{figure}
\centering
\includegraphics{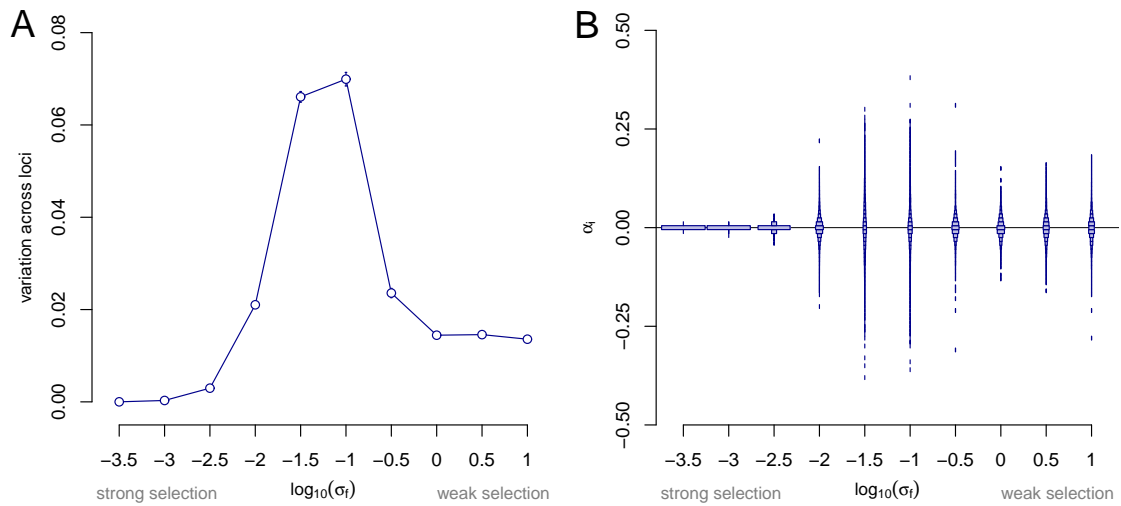}
	\caption{The distribution of direct effects of loci on the trait, $\alpha_i$, depends on the strength of selection on the trait, in the model without epistasis. A: Traits under intermediate selection (intermediate values of $\sigma_f$) have more variable effects. Under strong selection, the variance across loci is low because mutations changing the trait value are eliminated shortly and cannot be compensated by other mutations. Under weak selection, variance can increase through compensatory evolution when the architecture includes multiple loci, but this variance goes to $0$ when the number of loci reaches $1$. This occurs often enough (see the distributions in Fig. 1) to strongly reduce the mean variance of the phenotypic effects across loci. Parameters are set to their default values (table S2). B: This difference across traits under various strengths of selection is also apparent in the distribution of $\alpha_i$. The genetic architectures of traits under strong selection, and to a lesser extent of traits under weak selection are dominated by loci with small individual effects. Traits under intermediate selection rely on loci with more diverse contributions.}
	\label{fig:var}
\end{figure}

		\setkeys{Gin}{width=89mm}
	\begin{figure}
\centering
\includegraphics{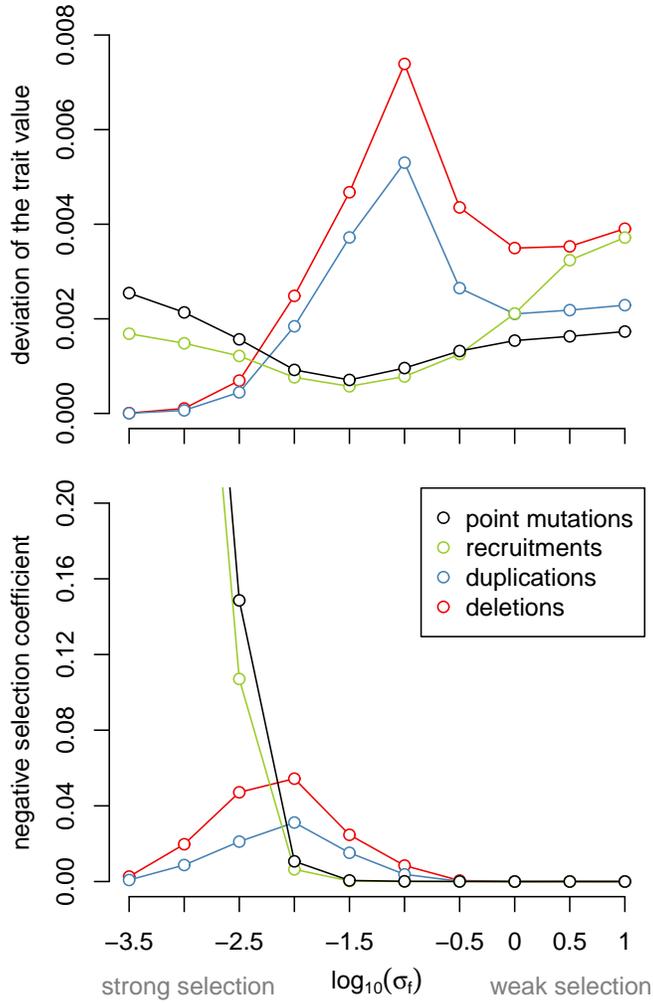}
	\caption{The effect of point mutations, duplications, deletions and recruitments on the trait value (top panel) is a non-monotonic function of the strength of selection on the trait. The effect of deletions and duplications on fitness (bottom panel) is also a non-monotonic function of the strength of selection on the trait, but the effect of point mutations and recruitments on fitness decreases continuously with $\sigma_f$. For each individual in the evolved populations of Fig. S3-A, we introduced $20$ mutations of each type and calculated the mean absolute effects on the trait and on fitness.}
	\label{fig:mutations}
\end{figure}

		\setkeys{Gin}{width=89mm}
	\begin{figure}
\centering
\includegraphics{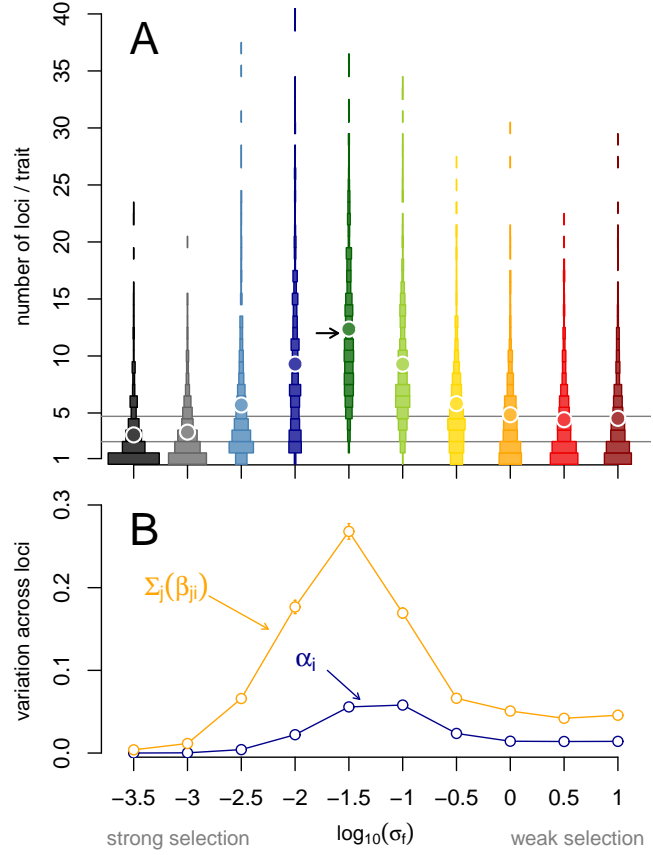}
	\caption{The genetic architecture underlying a trait depends on the strength of selection on the trait, in the presence of epistasis. A: Traits under intermediate selection (intermediate values of $\sigma_f$) evolve genetic architectures with the greatest number of controlling loci. The rectangle areas are proportional to the number of Wright-Fisher simulations (among $500$ per value of $\sigma_f$) in which the number of loci on the y-axis evolved. Dots denote the ensemble mean of each distribution. The neutral expectations for the equilibrium number of loci (see methods) are represented as grey lines, when recruitment events are neutral (top line) or not (bottom line; deletions and duplications are neutral in both cases). The black arrow represents the number of loci for which the number of deletions fixing approximately equals that of duplications or recruitments for $\sigma_f=10^{-1.5}$ (Fig. 2), where the mode of the distribution is expected. B: Standard deviations of $\alpha_i$ (direct effects) and $\sum_j (\beta_{ji})$ (indirect) are maximum under intermediate selection. Parameters are set to their default values (table S2).}
	\label{fig:architecture}
\end{figure}

		\setkeys{Gin}{width=150mm}
	\begin{figure}
\centering
\includegraphics{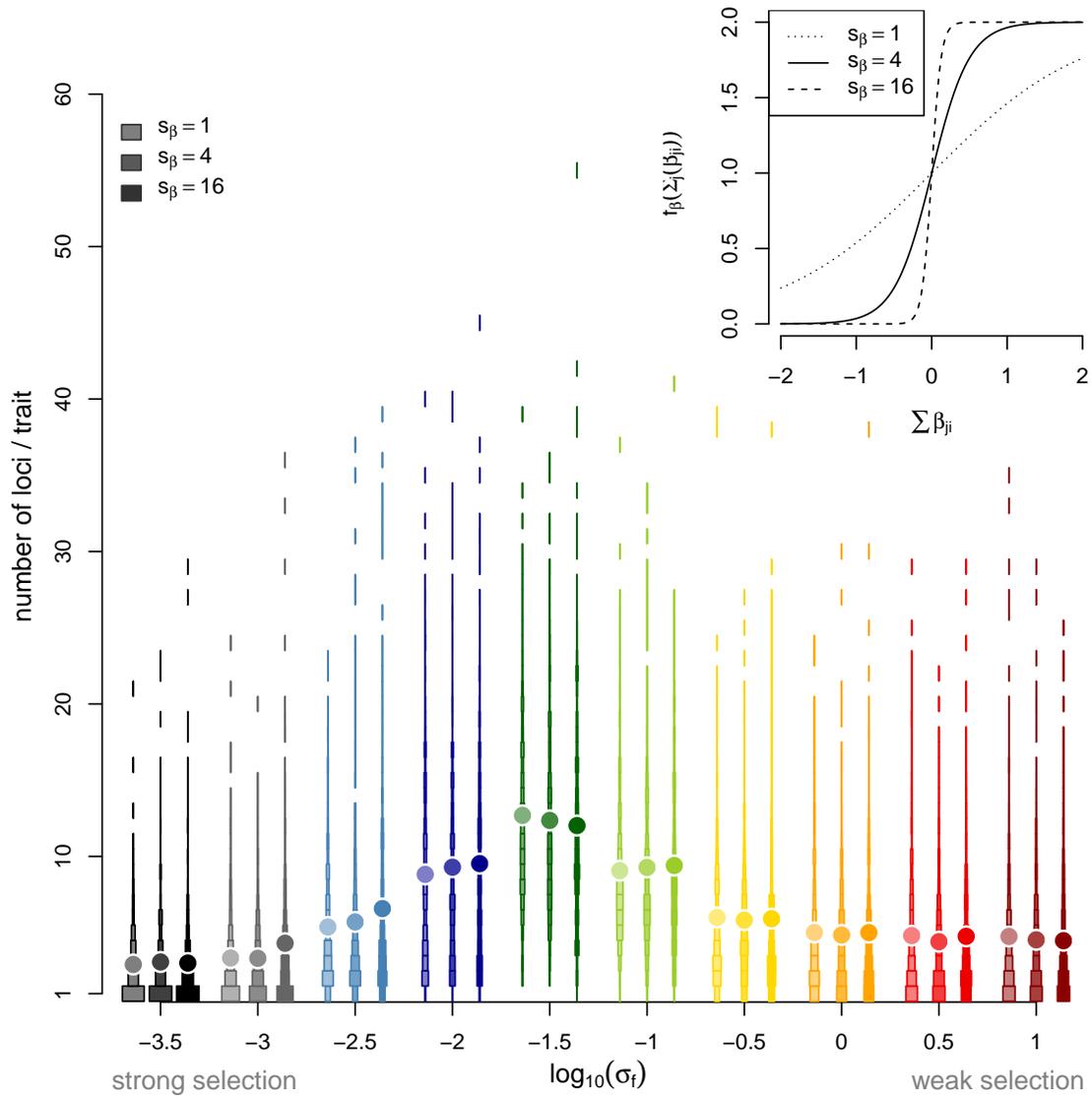}
	\caption{Impact of parameter $s_\beta$ on the evolution of genetic architecture. The insert on the top-right corner represents function $f_\beta$ for different values of $s_\beta$.}
	\label{fig:slope_beta}
\end{figure}

		\setkeys{Gin}{width=150mm}
	\begin{figure}
\centering
\includegraphics{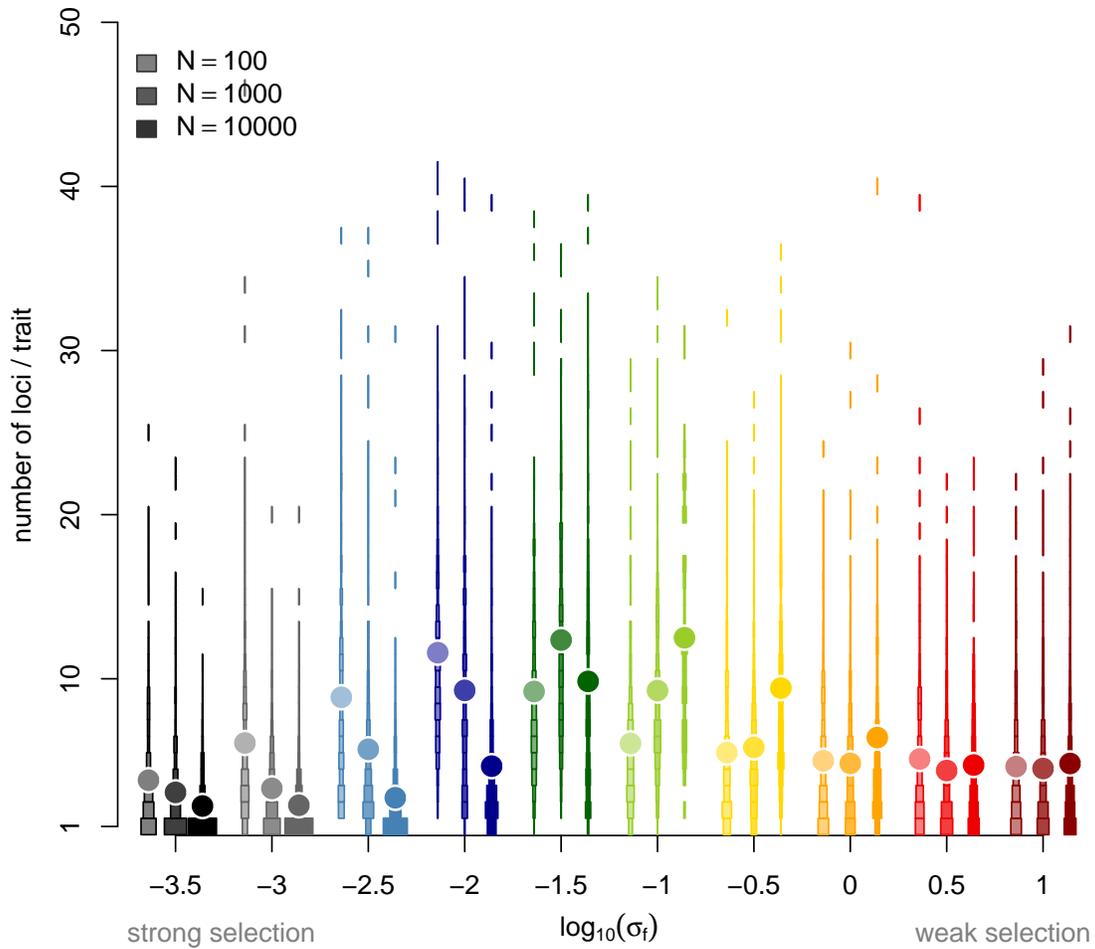}
	\caption{Increasing the population size increases the value of $\sigma_f$ at which the expected number of loci $L$ is maximum. All values represent the ensemble average of $500$ replicate simulations run for $5 \times 10^7$ generations.  All other parameters are set to their default values (table S2).}
	\label{fig:N}
\end{figure}

		\setkeys{Gin}{width=150mm}
	\begin{figure}
\centering
\includegraphics{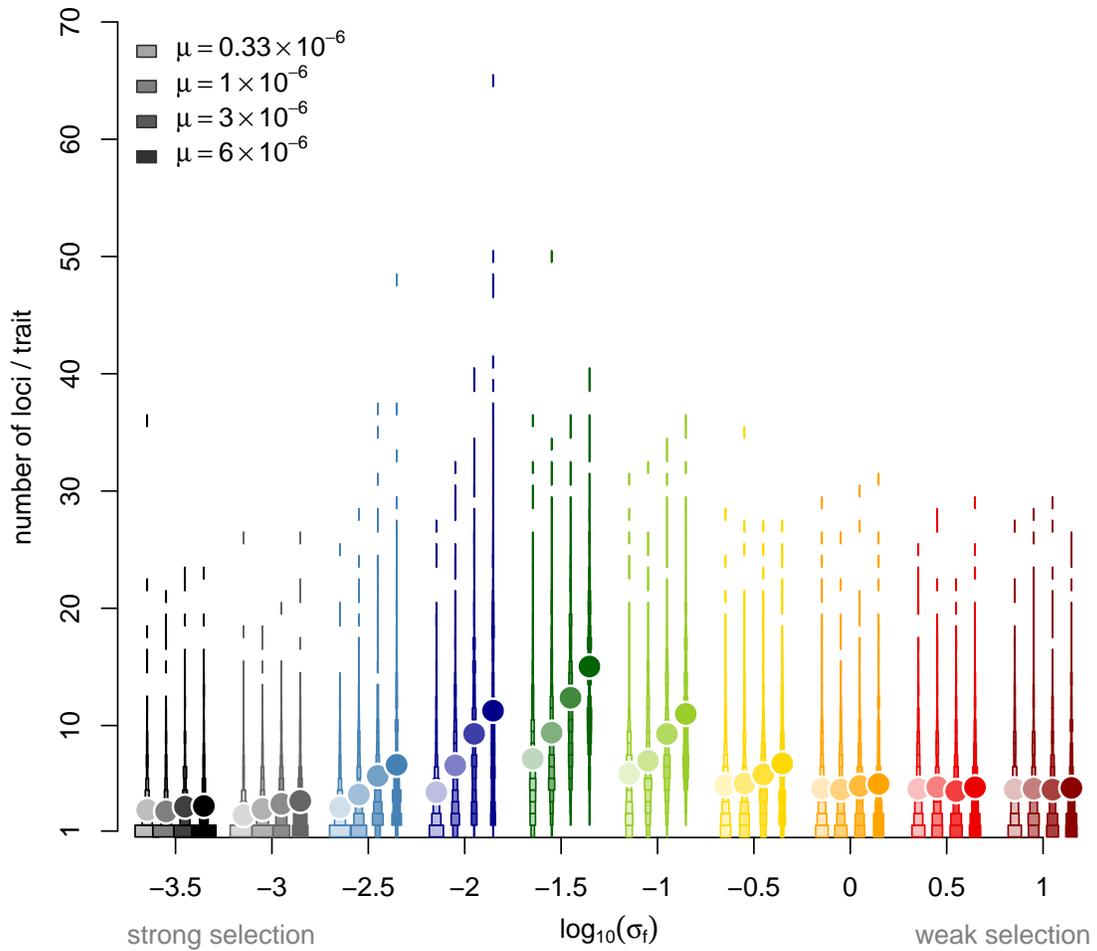}
	\caption{The expected number of loci $L$ increases with the mutation rate $\mu$. All values represent the ensemble average of $500$ replicate simulations run for $5 \times 10^7$ generations.  All other parameters are set to their default values (table S2). A mutation rate of $3 \times 10^{-6}$ was used to sample recruitment events, so the overall probability of recruitment remains constant.}
	\label{fig:mu}
\end{figure}

		\setkeys{Gin}{width=150mm}
	\begin{figure}
\centering
\includegraphics{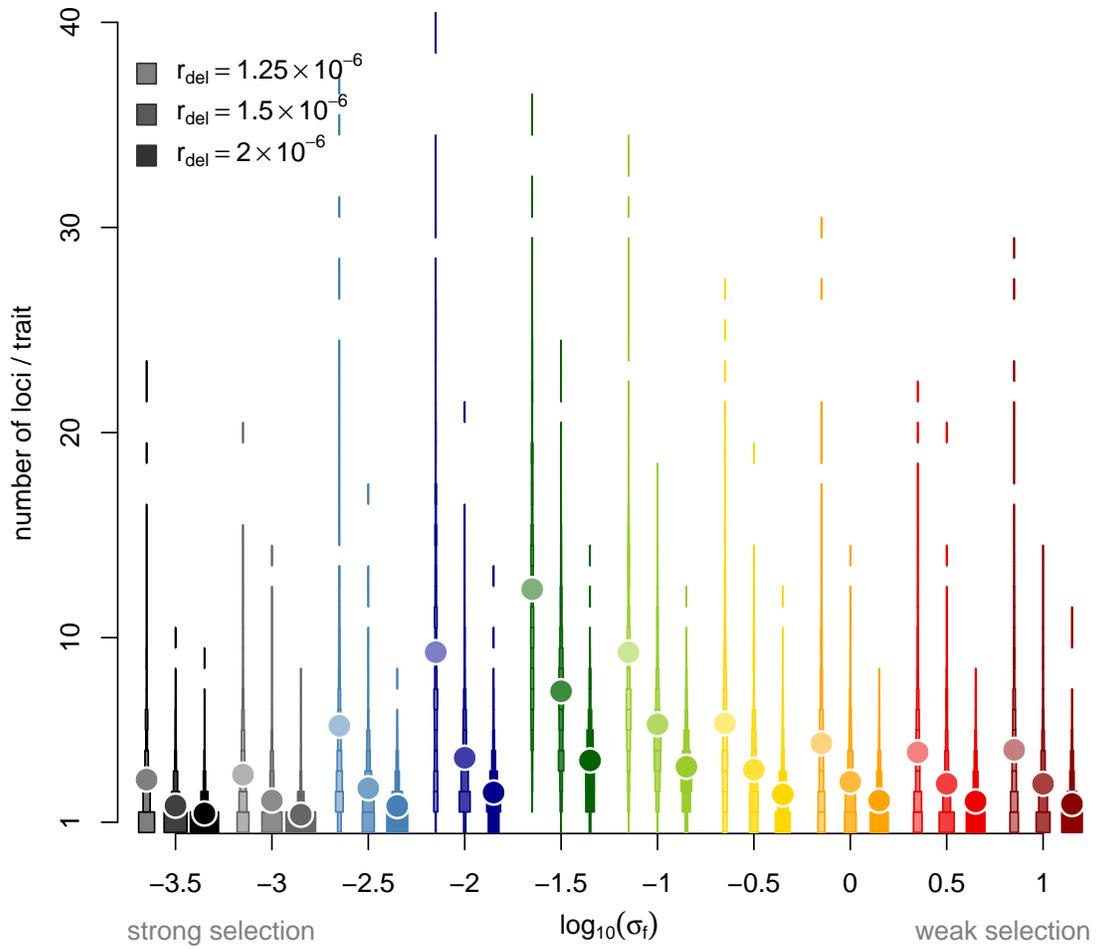}
	\caption{The expected number of loci $L$ decreases as deletions become more frequent (\textit{i.e.} $r_{del}$ increases). All values represent the ensemble average of $500$ replicate simulations run for $5 \times 10^7$ generations.  All other parameters are set to their default values (table S2).}
	\label{fig:r_del}
\end{figure}

		\setkeys{Gin}{width=150mm}
	\begin{figure}
\centering
\includegraphics{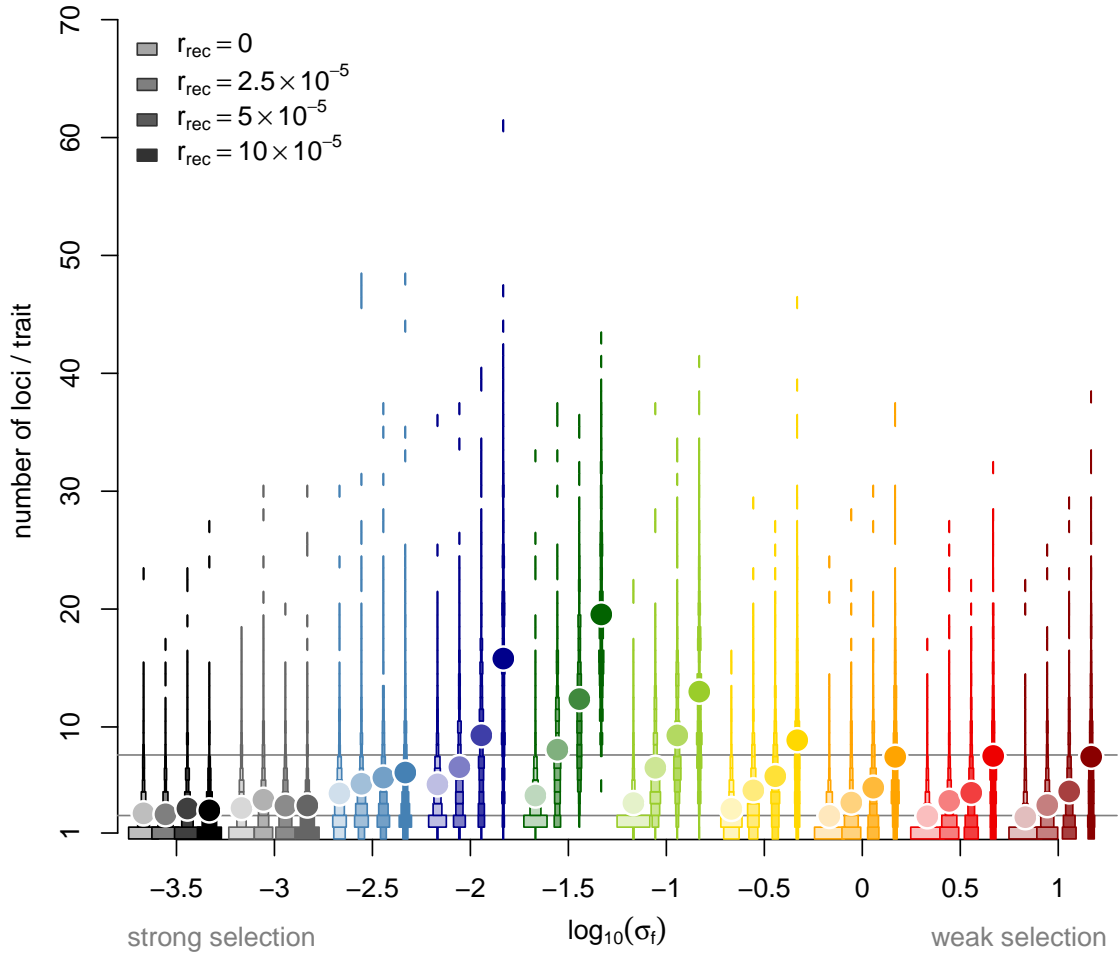}
	\caption{The expected number of loci $L$ increases as the probability of recruitment, $r_{rec}$, increases. All values represent the ensemble average of $500$ replicate simulations run for $5 \times 10^7$ generations.  All other parameters are set to their default values (table S2).}
	\label{fig:r_rec}
\end{figure}

		\setkeys{Gin}{width=150mm}
	\begin{figure}
\centering
\includegraphics{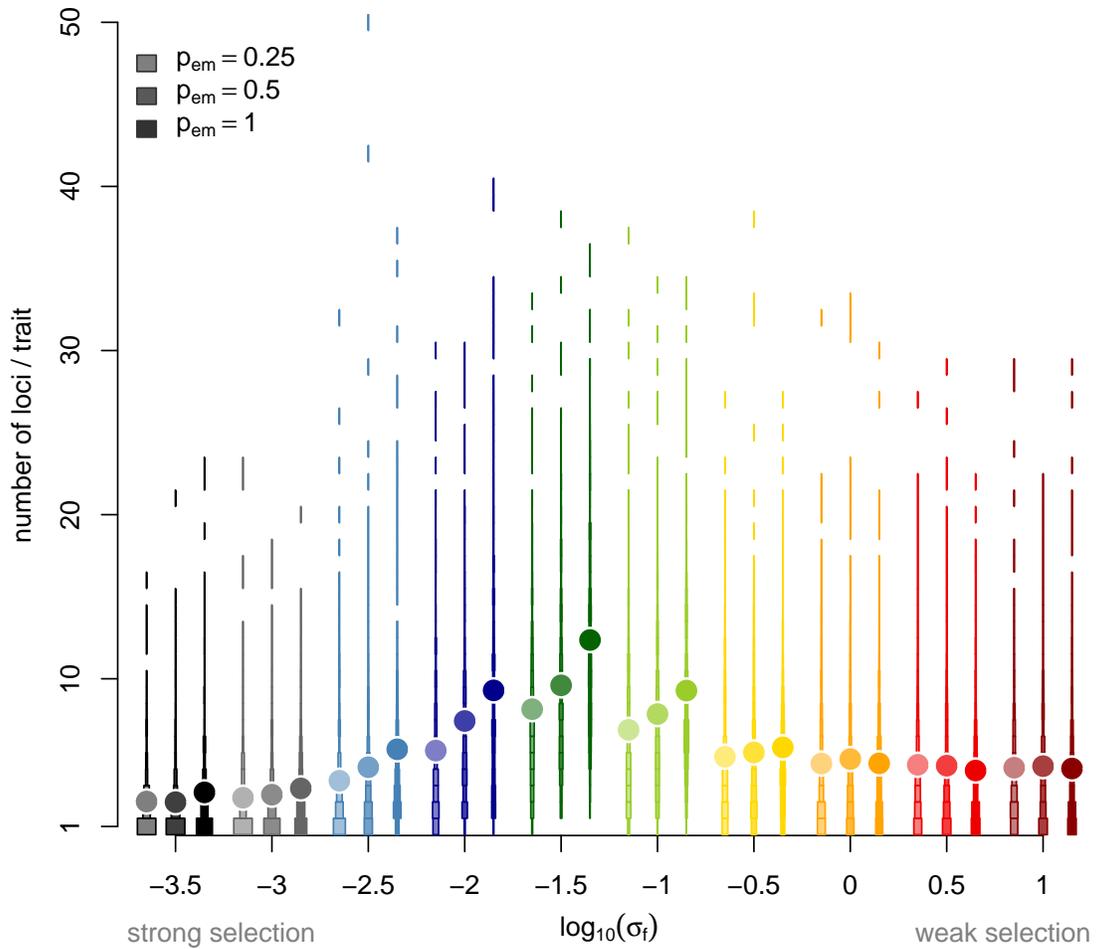}
	\caption{Decreasing the probability that a given subfunction is affected by a point mutation, $p_{em}$, has similar effects as decreasing $\mu$. All values represent the ensemble average of $500$ replicate simulations run for $5 \times 10^7$ generations.  All other parameters are set to their default values (table S2). }
	\label{fig:p_em}
\end{figure}

		\setkeys{Gin}{width=150mm}
	\begin{figure}
\centering
\includegraphics{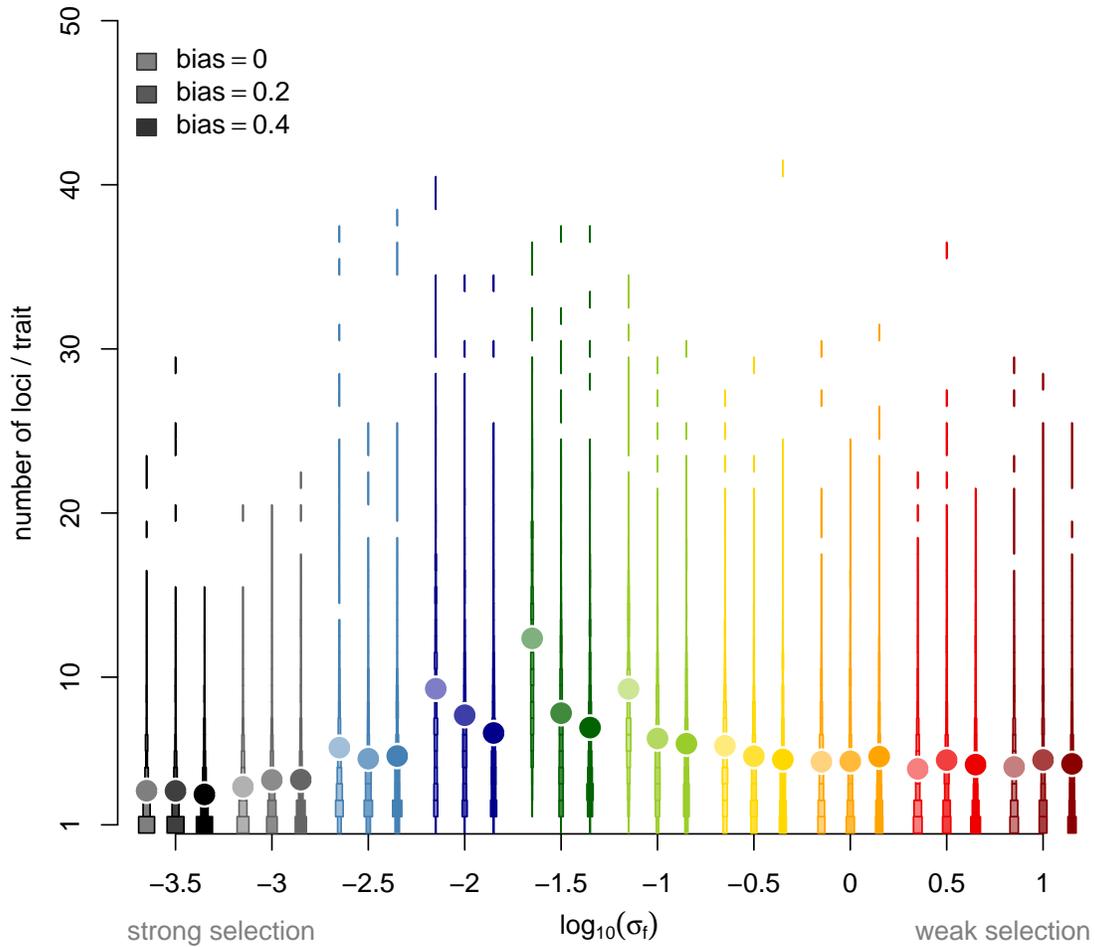}
	\caption{The evolution of $L$ is not strongly affected by mutation biases in $\alpha$ or $\beta$. A strong bias ($b_\alpha=b_\beta=0.4$) reduces the maximum variation across loci and therefore reduces $L$ when $\log_{10}(\sigma_f)>-2.5$.  All values represent the ensemble average of $500$ replicate simulations run for $5 \times 10^7$ generations. All other parameters are set to their default values (Table S2). One data point was omitted: $L \approx 61$ at $log_{10}(\sigma_f)=-1.5$ and $b_\alpha=b_\beta=0.2$. }
	\label{fig:bias}
\end{figure}

		\setkeys{Gin}{width=150mm}
	\begin{figure}
\centering
\includegraphics{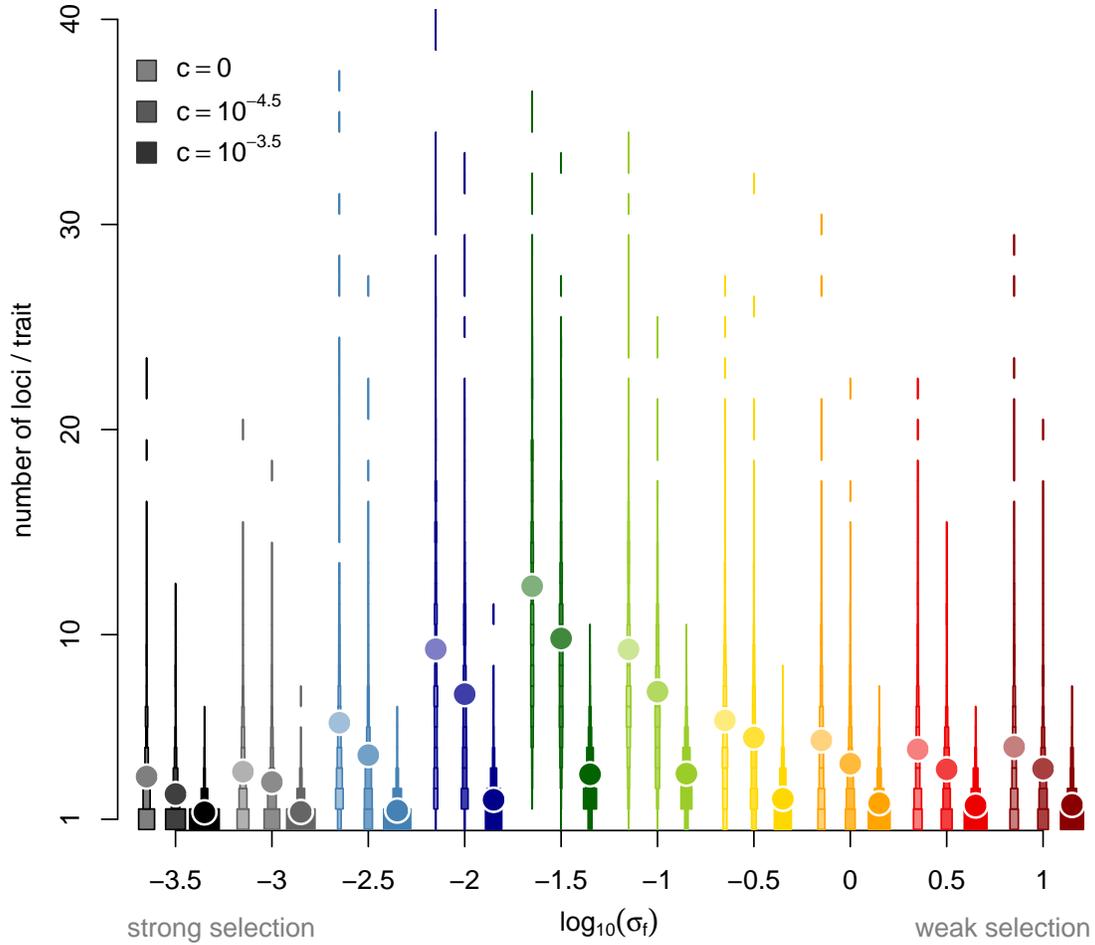}
	\caption{The expected value of $L$ decreases with the metabolic cost $c$. All values represent the ensemble average of $500$ replicate simulations run for $5 \times 10^7$ generations. All other parameters are set to their default values (table S2).}
	\label{fig:c}
\end{figure}

		\setkeys{Gin}{width=89mm}
	\begin{figure}
\centering
\includegraphics{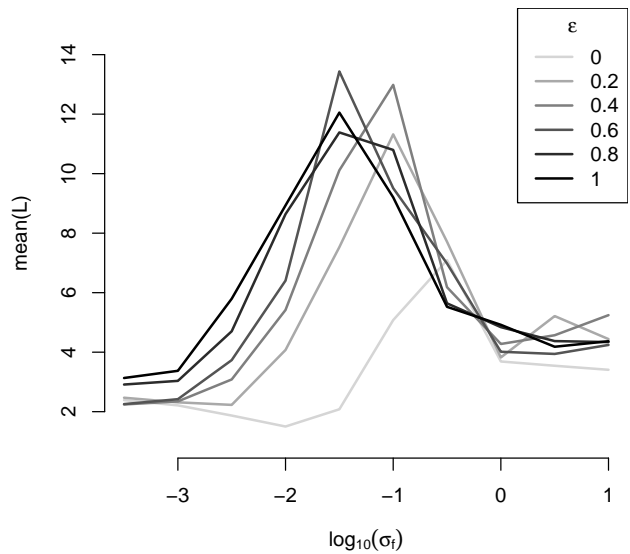}
	\caption{The number of loci contributing to a trait is a non-monotonic function of $\sigma_f$ whenever $\epsilon$ is higher than $0$ (Text S1). All values represent the ensemble average of $500$ replicate simulations run for $5 \times 10^7$ generations. All other parameters are set to their default values (table S2).}
	\label{fig:epsilon}
\end{figure}

		\setkeys{Gin}{width=150mm}
	\begin{figure}
\centering
\includegraphics{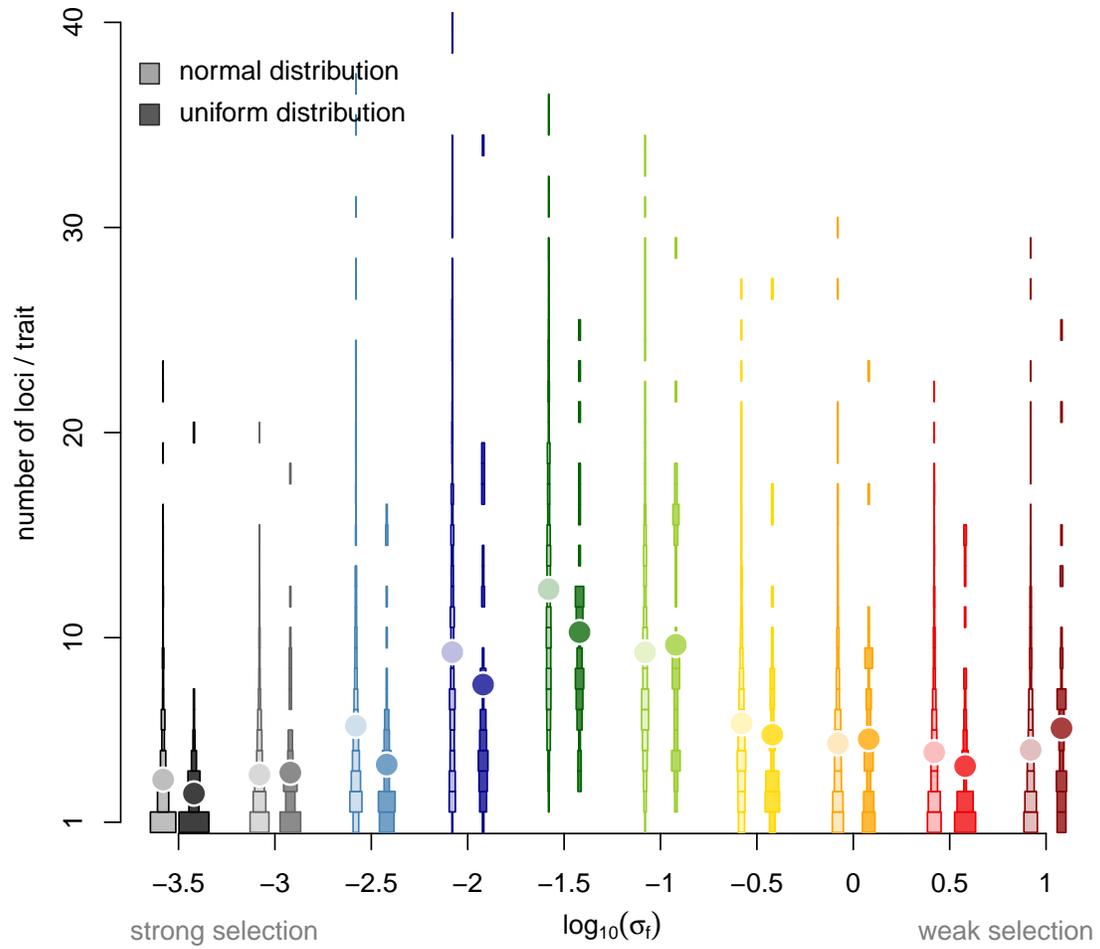}
	\caption{The expected value of $L$ is not affected by the form of the distribution of mutation effects. All values represent the ensemble average of $500$ replicate simulations run for $5 \times 10^7$ generations. All other parameters are set to their default values (table S2).}
	\label{fig:dist}
\end{figure}

		\setkeys{Gin}{width=89mm}
	\begin{figure}
\centering
\includegraphics{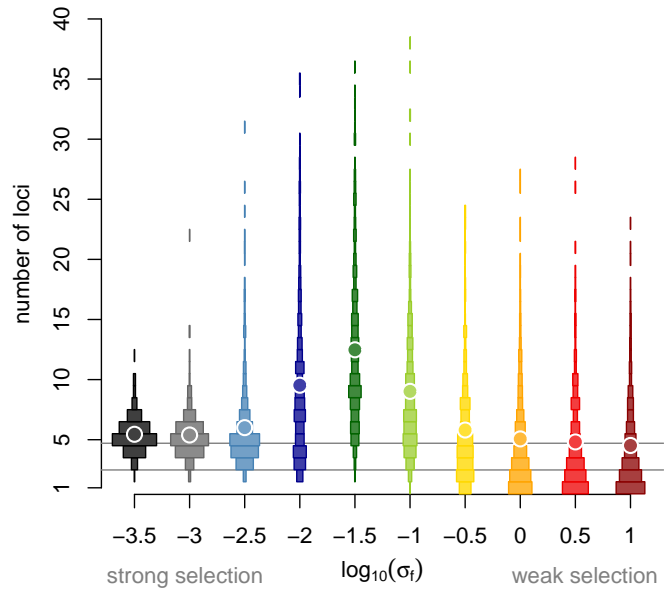}
	\caption{Evolution of the genetic architecture in simulations initiated with $5$ loci and variable effects (see description in the methods section). Compared to Fig. S3-A, only the architectures of traits under strong selection have changed. The mean number of loci under strong selection would be expected at the bottom grey line if deletions and duplications had neutral effects. Instead, this initial variation across loci prevents deletion or duplication, so the mean number of loci remains close to its initial value. Architectures of traits under intermediate and weak selection are not affected.}
	\label{fig:init}
\end{figure}

		\setkeys{Gin}{width=89mm}
	\begin{figure}
\centering
\includegraphics{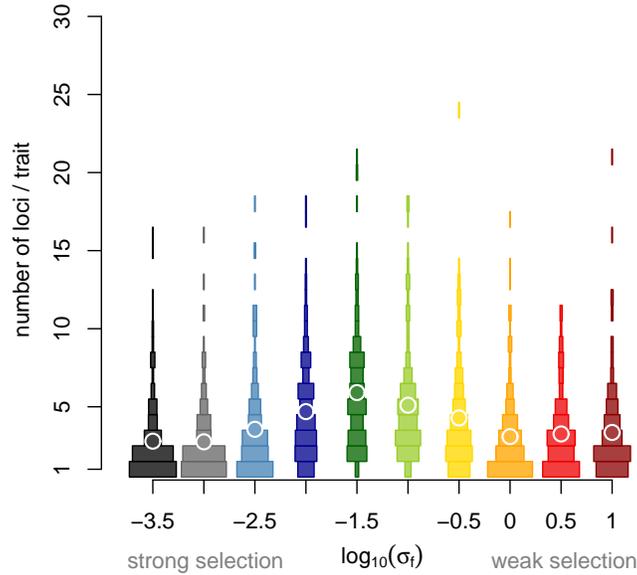}
	\caption{Evolution of genetic architectures in a multi-trait version of our model (see text S3). In this model the overall rate of recruitments of new loci is reduced, and therefore so too the equilibrium number of loci per trait. Nevertheless, the qualitative relationship between selection pressure and number of loci is similar to that in the single-locus version of our model.}
	\label{fig:multitrait}
\end{figure}

		\setkeys{Gin}{width=89mm}
	\begin{figure}
\centering
\includegraphics{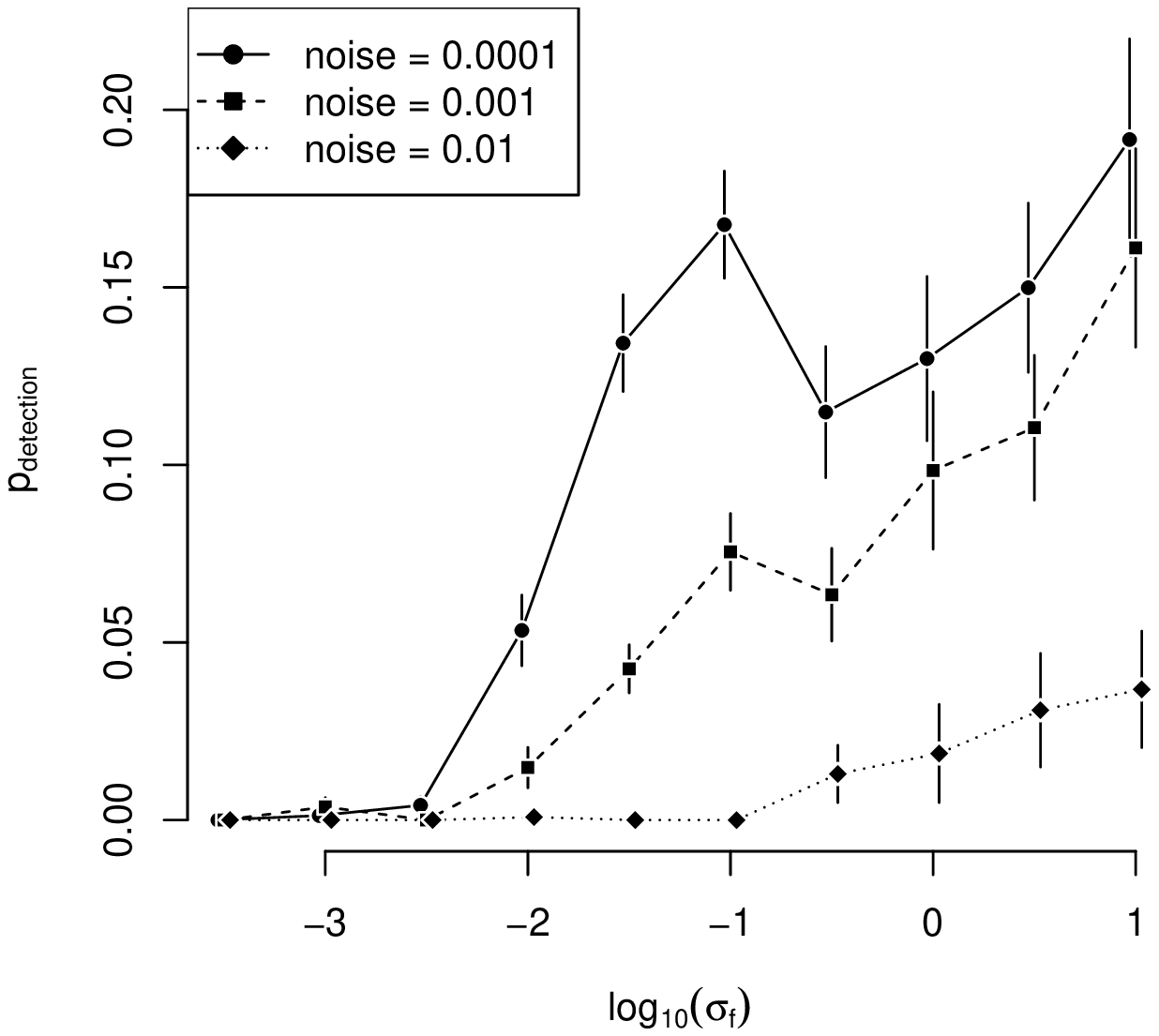}
	\caption{The probability to detect a locus in the true architecture increases as selection becomes weaker ($\sigma_f$ increase). Detection is more accurate as the noise decreases. Error bars represent the mean $\pm$ one standard error, calculated over $100$ replicate QTL simulations.}
	\label{fig:divergence}
\end{figure}

		\setkeys{Gin}{width=89mm}

	\begin{figure}
\centering
\includegraphics{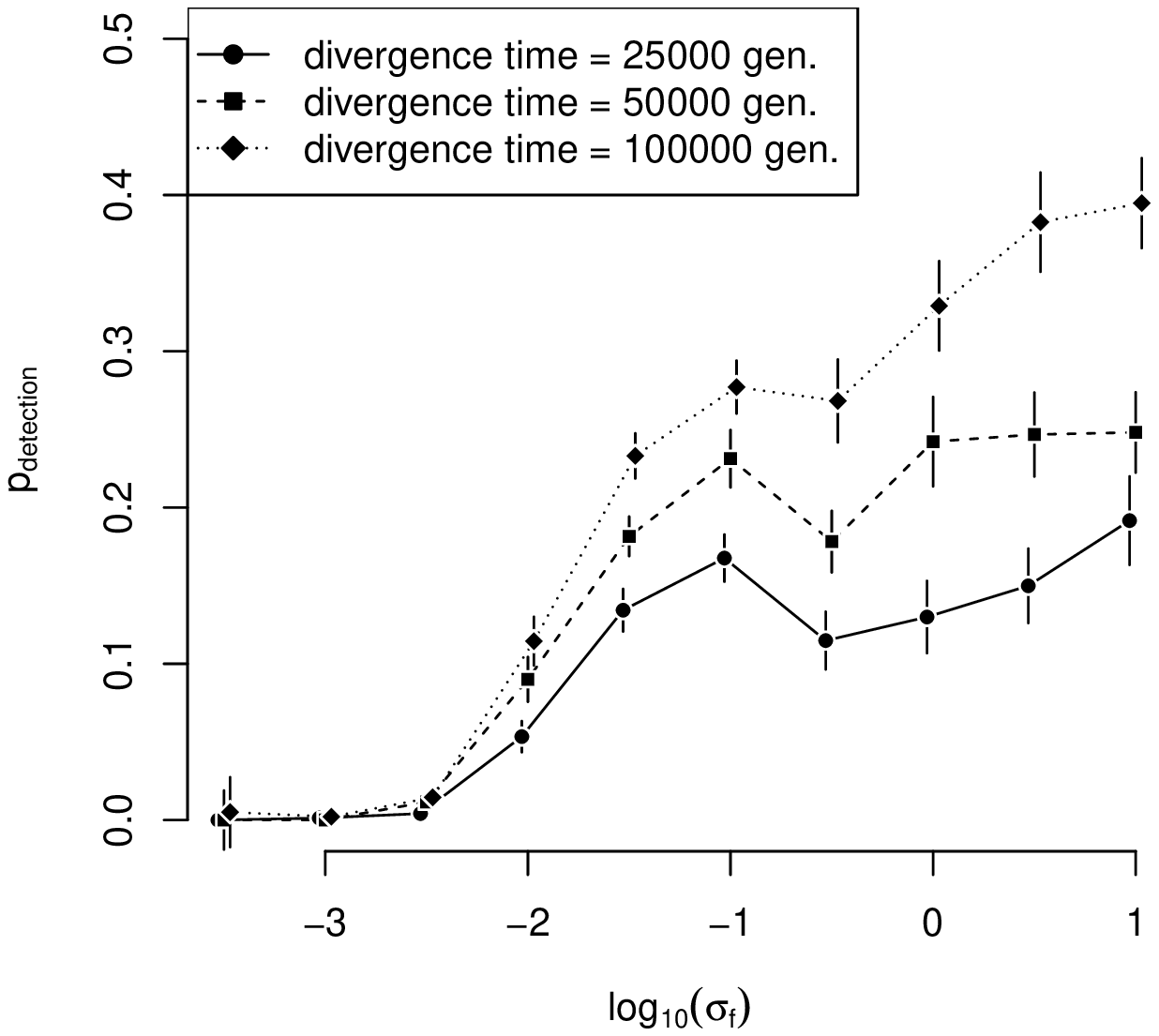}
	\caption{The probability to detect a locus in the true architecture increases as selection becomes weaker ($\sigma_f$ increase). Detection is more accurate as the divergence time increases. Error bars represent the mean $\pm$ one standard error, calculated over $100$ replicate QTL simulations.}
	\label{fig:divergence2}
\end{figure}

\clearpage
		\setkeys{Gin}{width=175mm}
	\begin{figure}
\centering
\includegraphics{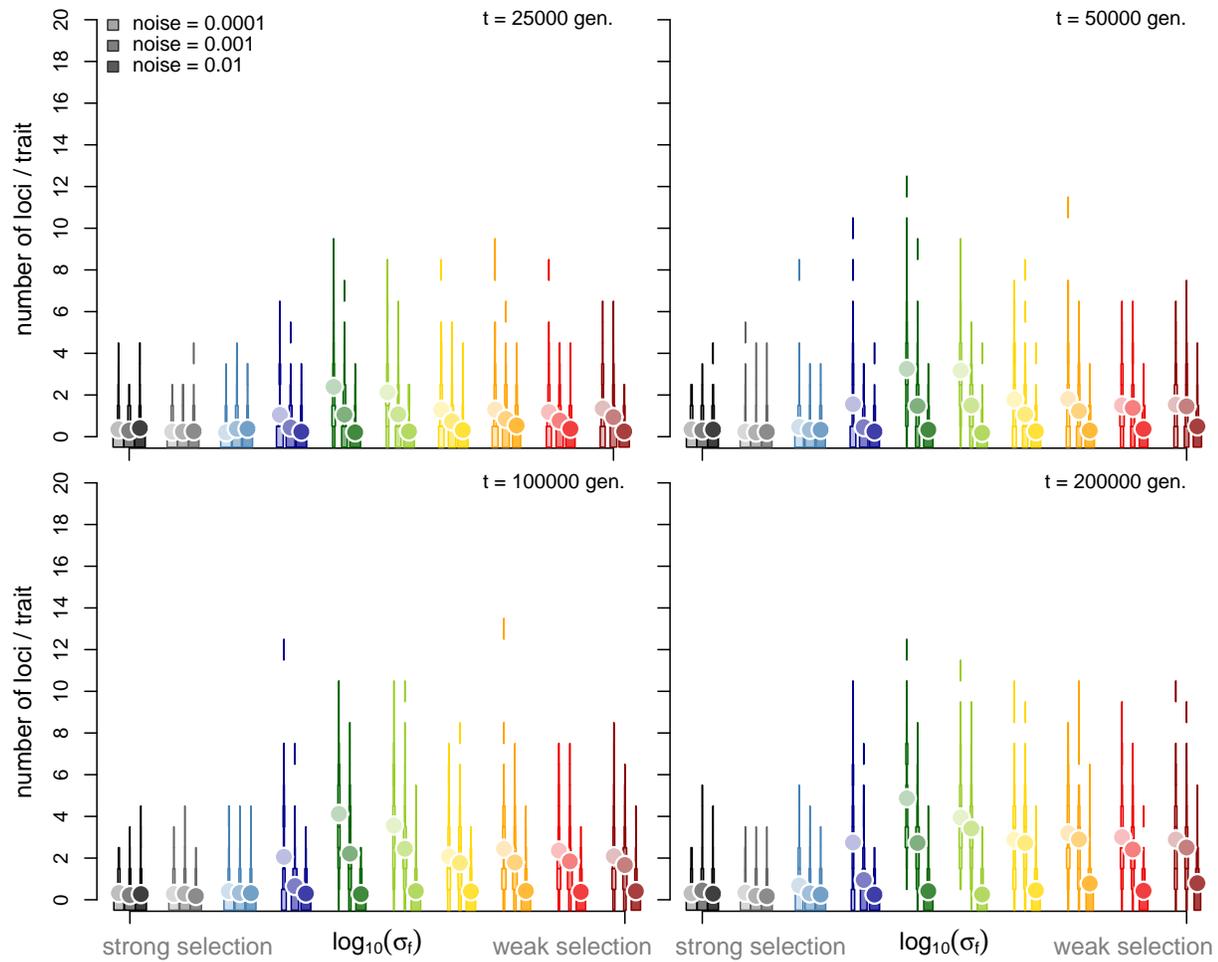}
	\caption{Number of QTL detected in a simulated study. The analysis is similar to Fig. 3B in the main text, but we changed the time of divergence between the two lines in the experiment (indicated in the top right corner of each panel) and the noise in traits measurements.}
	\label{fig:divergence3}
\end{figure}

		\setkeys{Gin}{width=150mm}
	\begin{figure}
\centering
\includegraphics{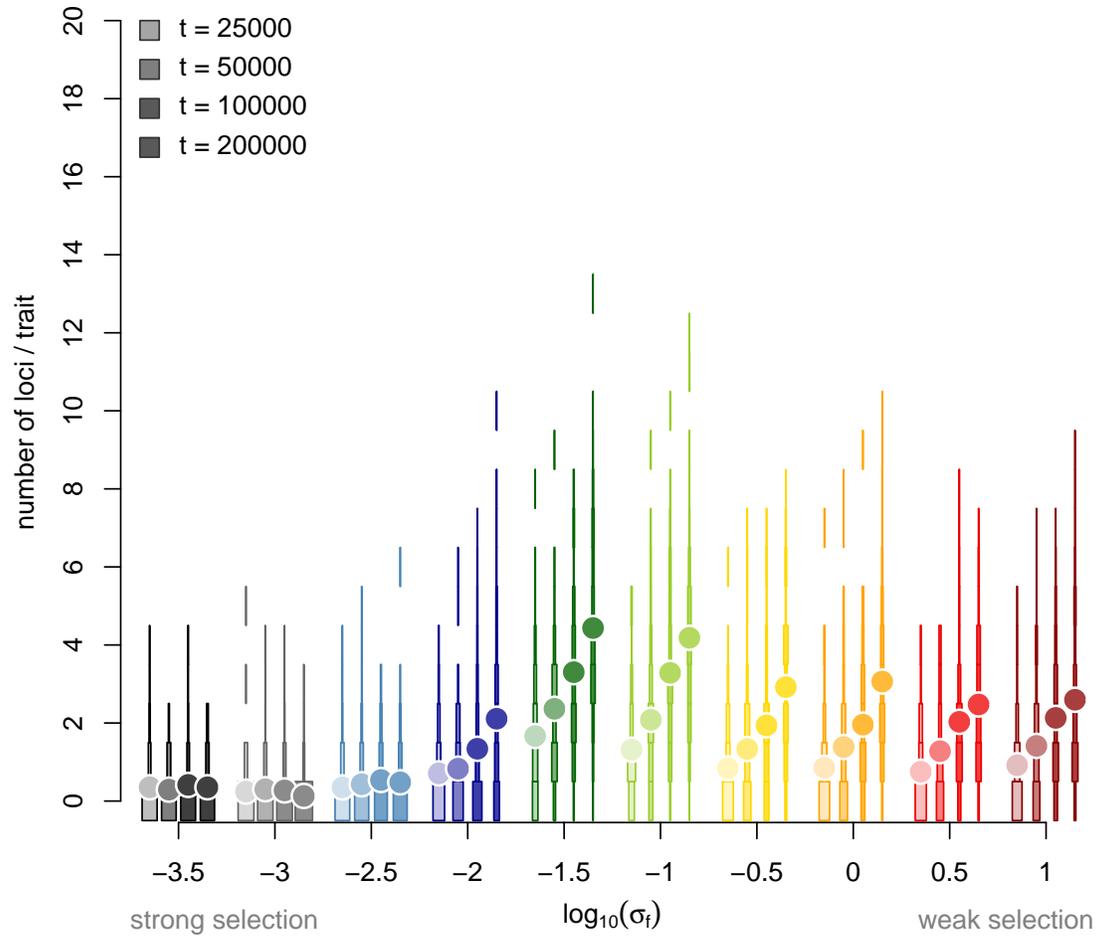}
	\caption{Number of QTL detected in a simulated study. The analysis is similar to Fig. 3B in the main text, but we changed the time of divergence between the two lines in the experiment. The noise in traits measurement increases proportionally to $\log_{10}(\sigma_f)$, from $0.0001$ to $0.001$.}
	\label{fig:divergence4}
\end{figure}		

\end{document}